\documentclass[a4paper,fleqn,usenatbib]{mnras}

\usepackage[T1]{fontenc}
\usepackage{ae,aecompl}

\usepackage{graphicx}	
\usepackage{amsmath}	
\usepackage{amssymb}	
\usepackage{natbib}
\usepackage{xspace}

\title[Detection of magnetic fields in K2 CP-stars]{Detection of magnetic
  fields in chemically peculiar stars observed with the K2 space mission}

\author[B.\,Buysschaert et al.]{
B.\,Buysschaert$^{1,2}$\thanks{E-mail: bram.buysschaert@obspm.fr},
C.\,Neiner$^{1}$,
A.\,J.\,Martin$^{1}$,
C.\,Aerts$^{2,3}$,
D.\,M.\,Bowman$^{2}$, \and
M.\,E.\,Oksala$^{4,1}$,
T.\,Van\,Reeth$^{2}$
\\
$^{1}$LESIA, Observatoire de Paris, PSL Research University, CNRS, Sorbonne Universit\'es, UPMC Univ. Paris 06,\\ Univ. Paris Diderot, Sorbonne Paris Cit\'e, 5 place Jules Janssen, F-92195 Meudon, France\\
$^{2}$Instituut voor Sterrenkunde, KU Leuven, Celestijnenlaan 200D, 3001 Leuven, Belgium\\
$^{3}$Dept. of Astrophysics, IMAPP, Radboud University Nijmegen, 6500 GL, Nijmegen, The Netherlands\\
$^{4}$Department of Physics, California Lutheran University, 60 West Olsen Road 3700, Thousand Oaks, CA, 91360, USA\\
}

\date{Accepted XXX. Received YYY; in original form ZZZ}

\pubyear{2017}

\begin{document}
\label{firstpage}
\pagerange{\pageref{firstpage}--\pageref{lastpage}}
\maketitle

\begin{abstract}
We report the results of an observational study aimed at searching for magnetic pulsating hot stars suitable for magneto-asteroseismology. A sample of sixteen chemically peculiar stars was selected and analysed using both high-resolution spectropolarimetry with ESPaDOnS and K2 high-precision space photometry.  For all stars, we derive the effective temperature, surface gravity,   rotational and non-rotational line broadening from our spectropolarimetric data. High-quality K2 light curves were obtained for thirteen of the sixteen stars and revealed rotational modulation, providing accurate rotation periods.  Two stars show evidence for roAp pulsations, and one star shows signatures of internal gravity waves or unresolved g-mode pulsations.  We confirm the presence of a large-scale magnetic field for eleven of the studied stars, of which nine are first detections. Further, we report one marginal detection and four non-detections. Two of the stars with a non-detected magnetic field show rotational modulation due to surface abundance inhomogeneities in the K2 light curve, and we confirm that the other two are chemically peculiar.  Thus, these five stars likely host a weak (undetected) large-scale magnetic field.
\end{abstract}

\begin{keywords}
stars: magnetic field -- stars: rotation -- stars: early-type -- stars: oscillations
\end{keywords}


\section{Introduction}

The number of known magnetic hot stars has steadily increased thanks dedicated campaigns (e.g., MiMeS, \citealt{2016MNRAS.456....2W}; the BOB campaign, \citealt{2015IAUS..307..342M}; and the BRITE spectropolarimetric survey, \citealt{2016arXiv161103285N}) using the latest generation of high-resolution spectropolarimeters.  Magnetic fields are found to occur in about 10\,\% of hot stars, are stable over long time scales, have a simple geometry (often dipolar), and are considered to be of fossil origin \citep[e.g.,][]{2007MsT..........1P, 2015IAUS..305...61N, 2016sf2a.conf..199V, 2017arXiv171202811S}.  Yet, we still do not understand the impact of such large-scale magnetic fields on the deep stellar interior or on stellar evolution.  Theory and numerical simulations indicate that the large-scale magnetic field should enforce a uniform rotation within the radiative envelope of hot stars \citep[e.g.,][]{1937MNRAS..97..458F, 1992MNRAS.257..593M, 1999A+A...349..189S, 2005A+A...440..653M, 2011IAUS..272...14Z}.  This affects the amount of material that is able to overshoot from the convective core to the surrounding radiative layers in these stars \citep[e.g.,][]{1981ApJ...245..286P, 2004ApJ...601..512B}.  However, observational evidence of this effect remains rare.

Asteroseismology, the study of non-radial oscillations in stars, remains the sole method to reliably observe the conditions within stars \citep{2010aste.book.....A}.  Depending on the type of stellar oscillations, classified on the basis of their dominant restoring force of the pulsation modes, the driving mechanism, and the global stellar parameters, these oscillations probe different regions within the stellar interior.  Gravity modes (g modes) probe the deep interior and the near-core region, while pressure modes (p modes) are mostly sensitive to the stellar envelope.  The detection and modelling of oscillation modes has been successfully exploited for only two hot magnetic OB stars, namely V2052\,Oph \citep{2012A+A...537A.148N, 2012MNRAS.424.2380H,   2012MNRAS.427..483B} and $\beta$\,Cep \citep{2000ApJ...531L.143S}, at the level of magneto-asteroseismology.  This is the combination of magnetometry and asteroseismology to study how the large-scale magnetic field of these magnetic hot stars influences their deep interior.  \citet{2012MNRAS.427..483B} determined that the magnetic field of V2052\,Oph was responsible for the low amount of convective-core overshooting required to successfully model the observed $\beta$\,Cep p-mode oscillations when, comparised to a star with similar oscillations but no detected large-scale magnetic field.

Unfortunately, the number of known magnetic pulsating hot stars with a complete characterization of their large-scale magnetic field and sufficient data to perform asteroseismic modelling remains limited.  This is in part due to the brightness limitation for ground-based high-resolution spectropolarimetry ($V = 7 - 10$\,mag is the typical limit for moderately rotating hot stars) and in part because few intrinsically bright hot stars were observed by photometric space missions, such as \textit{Kepler} or CoRoT.  The K2 space mission \citep{2014PASP..126..398H} has partly filled this niche, since its field of view changes on the sky every ${\sim}90$\,days, and monitors stars that are accessible with modern high-resolution spectropolarimetry.  As such, we are able to search for and investigate the periodic variability in photometry caused by oscillations and rotational modulation, while also measuring the strength of the large-scale magnetic field.

For this current study, we selected a sample of candidate magnetic hot stars, based on their peculiar chemical surface composition, to be observed with both K2 and ESPaDOnS, with the aim to find optimal targets for magneto-asteroseismology.

We describe the sample selection in Sect.\,\ref{sec:sample}.  Section\,\ref{sec:data} discusses the observations and their respective data treatments.  The periodic variability in the K2 photometry is investigated in Sect.\,\ref{sec:photometry}, estimates on the stellar parameters are determined in Sect.\,\ref{sec:spectroscopy}, and in Sect.\,\ref{sec:magnetic} we examine the high-resolution spectropolarimetry for the presence of a large-scale magnetic field.  Section\,\ref{sec:discussion} discusses the results, and a summary and conclusions are provided in Sect.\,\ref{sec:conclusions}.

\section{Sample selection}
\label{sec:sample}
For each observing campaign of the K2 space mission, we selected the CP-2 stars(also known as Ap/Bp stars) and the He weak/strong stars from the \citet{2009A+A...498..961R} catalogue, which contains a list of stars known to have peculiar chemical photospheric abundances.  The Ap/Bp stars are typically slow rotators with strong large-scale magnetic fields, which cause stratification in their atmospheres and create long-term chemical spots on the stellar surface \citep[e.g.,][]{1950MNRAS.110..395S, 1970ApJ...160..641M}.  The He strong stars are also assumed to host large-scale magnetic fields (e.g., \citealt{1979ApJ...228..809B}, with $\sigma$\,Ori\,E as the best studied example, \citealt{2015MNRAS.451.2015O}).  The He weak stars (CP-4 stars) are of a different nature and do not form a homogeneous group \citep[which was already noted by e.g.,][]{1964ARA+A...2..297S, 1971ApJS...23..213N, 1974VA.....16..131J}.  Some CP-4 stars exhibit intense Si, or Ti and Sr lines, and are considered a hot extension of the magnetic Ap/Bp stars (examples of magnetic He weak stars are 3\,Sco \citep{1979MNRAS.188..609L}, HD\,176582 \citep{2011AJ....141..169B}, and HR\,2949 \citep{2015MNRAS.449.3945S}).  Others show overabundances of P and Ga, typically noted for HgMn stars (CP-3 stars), and could thus be related to these non-magnetic HgMn stars.

In total, over 60 of these anticipated magnetic stars have been selected and monitored by the K2 space mission, in its campaigns C00 up to C15.  From this sample, we selected 16 of the brightest and/or slowest rotating stars to be observed with high-resolution spectropolarimetry, with the aim to detect and/or confirm the presence of a stable large-scale magnetic field.  The combined analysis of the photometric and spectro-polarimetric data of this sub-sample forms the basis   of the current paper.  The individual stars and their properties are presented in Table\,\ref{tab:sample}.  The photometric variability of the complete sample of 60 stars is analysed and presented in \citet[][submitted]{Bowman2018}.

Two stars of the sample are the primary components of long-period binary systems.  HD\,134759 is part of a hierarchical system, for which the shortest orbital period is 23.42\,yr \citep{1982A+AS...47..569H}.  HD\,158596 has an angular separation of ${\sim}0.3$\,arcsec according to its Hipparcos data \citep{1997ESASP1200.....E}.  One star is part of a short-period binary system, namely HD\,139160.  \citet{1987ApJS...64..487L} indicated an orbital orbit of 5.28\,d and an eccentricity of $0.33 \pm 0.02$.  Yet, substantial scatter remains between their spectroscopic observations and orbital solution.

\begin{table*}
\caption{Ap/Bp stars observed with K2 and ESPaDOnS.  Spectral types are retrieved from \citet{2009A+A...498..961R} and confirmed binary systems in the literature are indicated (see text).  Values for the effective temperature $T_{\mathrm{eff}}$ and surface gravity $\log g$ were determined in this work by fitting synthetic spectra to the H$\beta$ and H$\gamma$ lines.  The values for the projected rotational velocity $v\sin i$ and the non-rotational line broadening $v_{\rm NR}$ were determined from a fit to an unblended metal line.}
\centering
\tabcolsep=6pt
\begin{tabular}{lllrrrr}
\hline
\hline
Star		& V 		& SpT		& $T_{\mathrm{eff}}	$& $\log g	$& $v\sin i$& $v_{\rm NR}$\\
		& [mag]	&			& [K]				& [dex] 		& [km s$\mathrm{^{-1}}$]& [km s$\mathrm{^{-1}}$]\\
\hline
HD97859	&	9.35	&	B9 Si		&$	13750	\pm	300	$&$	3.85	\pm	0.10	$&$	83	\pm	1	$&$	42	\pm	5	$\\
HD107000	&	8.02	&	A2 Sr		&$	7850		\pm	200	$&$	3.20	\pm	0.15	$&$	20	\pm	5	$&$	41	\pm	6	$\\
HD134759	&	4.54	&	Bp Si + ...	&$	11900	\pm	200	$&$	3.80	\pm	0.10	$&$	60	\pm	2	$&$	33	\pm	4	$\\
HD139160	&	6.19	&	B7 He wk.	&$	13200	\pm	400	$&$	4.05	\pm	0.10	$&$	20	\pm	5	$&$	35	\pm	5	$\\
HD152366	&	8.08	&	B8 Si		&$	10250	\pm	450	$&$	3.30	\pm	0.30	$&$	23	\pm	2	$&$	17	\pm	3	$\\
HD152834	&	8.83	&	A0 Si		&$	10100	\pm	500	$&$	3.40	\pm	0.30	$&$	13	\pm	1	$&$	10	\pm	2	$\\
HD155127	&	8.38	&	B9 Eu Cr Sr	&$	8050		\pm	150	$&$	3.45	\pm	0.15	$&$	37	\pm	3	$&$	35	\pm	5	$\\
HD158596	&	8.94	&	B9 Si + ...	&$	11200	\pm	400	$&$	3.90	\pm	0.20	$&$	60	\pm	3	$&$	40	\pm	4	$\\
HD164224	&	8.49	&	B9 Eu Cr		&$	8850		\pm	250	$&$	3.45	\pm	0.10	$&$	22	\pm	4	$&$	40	\pm	7	$\\
HD165972	&	8.96	&	B9 Si		&$	12700	\pm	250	$&$	3.85	\pm	0.10	$&$	29	\pm	4	$&$	50	\pm	5	$\\
HD166804	&	8.88	&	B9 Si		&$	12900	\pm	250	$&$	4.00	\pm	0.10	$&$	45	\pm	3	$&$	29	\pm	4	$\\
HD173406	&	7.43	&	B9 Si		&$	13150	\pm	250	$&$	4.05	\pm	0.10	$&$	38	\pm	2	$&$	33	\pm	5	$\\
HD173657	&	7.41	&	B9 Si Cr		&$	10750	\pm	500	$&$	3.80	\pm	0.20	$&$	91	\pm	5	$&$	47	\pm	15	$\\
HD177013	&	9.04	&	A2 Eu Cr Sr	&$	11700	\pm	400	$&$	4.40	\pm	0.20	$&$	24	\pm	6	$&$	41	\pm	8	$\\
HD177562	&	7.38	&	B8 Si		&$	12200	\pm	300	$&$	3.65	\pm	0.10	$&$	15	\pm	1	$&$	13	\pm	3	$\\
HD177765	&	9.15	&	A5 Eu Cr Sr	&$	8600		\pm	200	$&$	4.50	\pm	0.10	$&$	5	\pm	2	$&$	8	\pm	2	$\\

\hline
\end{tabular}
\label{tab:sample}
\end{table*}

\subsection{Previous inference of large-scale magnetic fields}
For some of the stars in our sample, the presence of a large-scale magnetic field has been investigated from previous observations.  However, in several cases it was not clearly confirmed that the star hosts a large-scale magnetic field.  These observations are either from a photometric estimation of the `surface' magnetic field \citep[e.g.,][]{1980A+AS...41..111C}, medium resolution circular spectropolarimetry of metal or Balmer lines \citep[e.g.,][]{2014AstBu..69..427R}, or from the directly observed Zeeman splitting of absorption lines \citep{1997A+AS..123..353M}.

The photometric $Z$ index based on the Geneva photometric system and calibrated on Kurucz models \citep{1979ApJS...40....1K} can indicate the presence of a (large-scale) magnetic field for hot stars \citep[with a spectral type ranging from B1 to A2;][]{1979A+A....78..305C, 1980A+A....88..135C}.  This method was employed by \citet{1980A+AS...41..111C} to indicate that HD\,134759 and HD\,139160 should host a large-scale magnetic field with a strength of more than 1\,kG.  \citet{1984A+AS...58..387N} continued this analysis and determined a magnetic field strength of 1\,kG, 2.7\,kG, 4.1\,kG, 2.4\,kG, and 1.2\,kG for HD\,134759, HD\,155127, HD\,164224, HD\,165972, and HD\,173406, respectively.

Low to medium resolution circular spectropolarimetry was utilized to detect and measure the strength of the large-scale magnetic fields of three stars in our sample.  \citet{1975ApJ...201..624L} investigated HD\,134759 and obtained a longitudinal magnetic field value of $-600\pm 700$\,G, which the authors did not claim as a definite detection.  HD\,97859 was observed and discussed by \citet{1998CoSka..27..452E}, who obtained a longitudinal magnetic field of $-400$\,G.  No errorbars were provided, but it can be assumed to be of the order of the measured value, as HD\,97859 had the largest $v \sin i$ of all stars in their sample, with typical uncertainties ranging from 100\,G to 500\,G for stars with a $v \sin i < 20$\,km\,s$\mathrm{^{-1}}$.  Lastly, HD\,107000 was monitored by \citet{2008AstBu..63..139R, 2014AstBu..69..427R, 2015AstBu..70..444R}, who measured longitudinal magnetic field values ranging from $-240\pm30$\,G up to $320\pm50$\,G that vary with a period of about 2.4\,d.

\citet{1997A+AS..123..353M} and \citet{2017A+A...601A..14M} noted that HD\,177765 hosts a sufficiently strong large-scale magnetic field causing Zeeman splitting of several absorption lines.  By measuring the strength of the splitting, the authors determined a mean magnetic field modulus of approximately 3.4\,kG, with little variation between the various observations.

Given this overview, we conclude that only HD\,107000 and HD\,177765 have a firmly detected large-scale magnetic field, albeit not with modern high-resolution spectropolarimetry.  

\section{Observations}
\label{sec:data}
\subsection{K2 photometry}

To construct the K2 light curves, we retrieved the target pixel files from the Mikulski Archive for Space Telescopes (MAST).  For each (sub)-campaign and each star, we determined custom non-circular apertures from the stacked images.  These were sufficiently large to accommodate the pixel drifts due to the thruster fires \citep{2014PASP..126..398H} and did not change during a given data set.  Next, the mean background per frame was subtracted to create raw background corrected photometry.  This was subsequently passed through the \textsc{k2sc} package \citep{2015MNRAS.447.2880A, 2016MNRAS.459.2408A, 2017MNRAS.471..759A} to correct for the pixel drifts, and their associated instrumental effects, by means of Gaussian processes.  Subsequently, an outlier rejection was applied to the corrected photometry of each (sub)-campaign and another (long-term) instrumental detrending was performed.  During this detrending step, we accounted for the high-amplitude rotational modulation signal (see Sect.\,\ref{sec:photometry_rotation}).  Finally, we combined the different (sub)-campaigns for a given star to one light curve.  Details on the K2 photometry are given in Table\,\ref{tab:photometry}.

Reduced K2 light curves were constructed for 14 stars of the sample, while reductions for HD\,134759 and HD\,139160 were still ongoing at the time of writing.  A significant fraction of the flux of HD\,177562 was lost outside the subraster, causing the light curve to be of poor quality. Therefore, only 13 light curves were used in the remaining of this paper.

\begin{table}
\caption{Logs of the the K2 photometric observations.  We provide the EPIC ID, the K2 campaign number, C, during which observations were taken, the total duration of the observations, and the total number of datapoints in the light curve, N.  $^{a}$ Light curve not yet fully reduced at the time of submission.  $^{b}$ Light curve has a significant time gap in between the two campaigns.  $^{c}$ Light curve of poor quality due to flux loss outside the sub-raster.}
\centering
\tabcolsep=6pt
\resizebox{\columnwidth}{!}{
\begin{tabular}{lllccc}
\hline
\hline
Star		& EPIC ID		& C& Time length	& N & Notes\\
		&&&[d]\\
\hline
HD\,97859	&	201777342	&	01	&	80.07	&	3345	&$		$\\
HD\,107000	&	201667495	&	10	&	47.89	&	1907	&$		$\\
HD\,134759	&	249657024	&	15	&			&		&$	^{a}	$\\
HD\,139160	&	249152551	&	15	&			&		&$	^{a}	$\\
HD\,152366	&	203749199	&	02, 11&	822.59	&	4209	&$	^{b}	$\\
HD\,152834	&	232176043	&	11	&	69.43	&	2925	&$		$\\
HD\,155127	&	232284277	&	11	&	69.45	&	2876	&$		$\\
HD\,158596	&	225990054	&	09, 11&	221.99	&	5831	&$	^{b}	$\\
HD\,164224	&	226241087	&	09	&	68.64	&	2824	&$		$\\
HD\,165972	&	224206658	&	09	&	68.64	&	2822	&$		$\\
HD\,166804	&	227373493	&	09	&	68.64	&	2812	&$		$\\
HD\,173406	&	218676652	&	07	&	65.79	&	2624	&$		$\\
HD\,173657	&	213786701	&	07	&	81.28	&	3361	&$		$\\
HD\,177013	&	219198038	&	07	&	81.24	&	3289	&$		$\\
HD\,177562	&	214133870	&	07	&	81.80	&	3494	&$	^{c}	$\\
HD\,177765	&	214503319	&	07	&	78.78	&	3331	&$		$\\
\hline
\end{tabular}}
\label{tab:photometry}
\end{table}

\subsection{ESPaDOnS spectropolarimetry}

Each star was observed at least once by the Echelle SpectroPolarimetric Device for the Observations of Stars \citep[ESPaDOnS,][]{2006ASPC..358..362D} mounted at the Canada France Hawaii Telescope (CFHT) on Mauna Kea in Hawaii.  Standard settings were used for the spectropolarimeter in circular polarisation mode (Stokes\,V).  The spectropolarimetric sequences consisted of four consecutive sub-exposures, for which the exposure times were tailored to the spectral type, brightness, and anticipated $v\sin i$, to be able to detect a dipolar field of 600\,G or higher.  The observations were reduced with the \textsc{libre-esprit} \citep{1997MNRAS.291..658D} and \textsc{upena} softwares available at CFHT.  The data span from 3\,700\,\AA\ to 10\,500\,\AA\ and have an average resolving power R\,=\,65\,000.  The spectropolarimetric data were normalized to unity per spectral order with the interactive spline fitting tool \textsc{spent} \citep{2018MNRAS.475.1521M}.  Details on the spectropolarimetric observations are given in Table\,\ref{tab:magnetometry}.

\section{Photometric variability}
\label{sec:photometry}

\begin{figure*}
\centering
\includegraphics[width=\textwidth, height = 0.33\textheight]{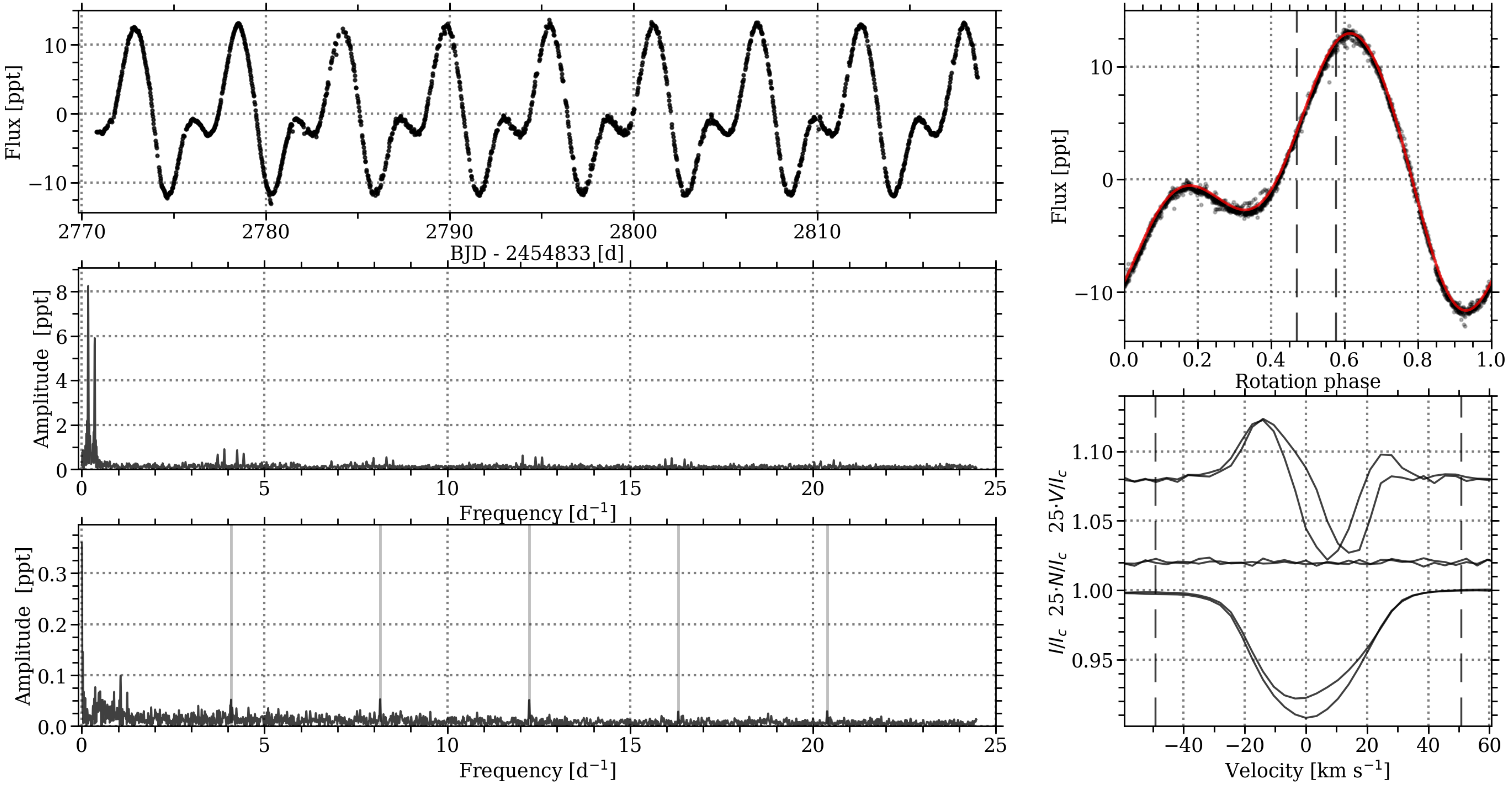}%
\caption{The K2 light curve of HD\,107000 is shown in the \textit{top left} panel and its corresponding Lomb-Scargle periodogram in the \textit{middle left} panel.  \textit{Top right}: K2 photometry (black) phase-folded with the rotation period, as well as a fourth-order sine model (red).  Rotation phases where ESPaDOnS spectropolarimetry were taken are marked by the black dashed lines.  \textit{Bottom left}: Lomb-Scargle periodogram of the residual K2 photometry, after removal of the rotational modulation signal.  Instrumental frequencies related to the thruster firing are indicated by the gray solid lines.  \textit{Bottom right}:  LSD profiles of the ESPaDOnS spectropolarimetry where the Stokes\,V, the diagnostic null, and the Stokes\,I profiles are shown, with an offset for increased visibility.  Integration limits for the determination of the FAP and longitudinal magnetic field are indicated by the dashed vertical lines.}
\label{fig:hd107000}
\end{figure*}

\subsection{Rotational modulation}
\label{sec:photometry_rotation}
We computed a Lomb-Scargle periodogram \citep{1976Ap+SS..39..447L, 1982ApJ...263..835S} for each of the 13 K2 light curves.  The dominant variability for all 13 light curves occurs in the low-frequency regime, which consists of a few significant frequency harmonics.  These periodograms are indicated in the top and middle panels of the left column of Fig.\,\ref{fig:hd107000} (similar figures for the other stars are depicted in Appendix\,\ref{sec:appendix}).  We attribute this variability to rotational modulation, due to surface abundance inhomogeneities.  Instead of iteratively prewhitening this variability, we performed a Least-Squares (LS) fit to the light curve with a fourth-order sine model:
\begin{equation}
M(t) = C + \sum\limits_{n=1}^{4} A_n \cdot \sin \left(2 \pi (n f_{\mathrm{rot}} t + \phi_n)\right) \, \mathrm{,}
\label{eq:rotmodel}
\end{equation}
\noindent where $A_n$ and $\phi_n$ are the amplitude and phase of the $n$th harmonic and $C$ is a constant offset.  This fitting procedure has the advantage that the exact rotation frequency, $f_{\rm rot}$, and its (higher) frequency harmonics are simultaneously enforced.  We truncated the series at the fourth order, because this was the highest frequency harmonic encountered in the periodograms.  The fitting process was a two-step approach.  First, $f_{\mathrm{rot}}$ was kept fixed to the estimated value from a ten times oversampled Lomb-Scargle periodogram to estimate $A_n$, $\phi_n$, and $C$.  In the second step, all parameters, including $f_{\mathrm{rot}}$, were set as free fitting parameters, where $f_{\rm rot}$ was allowed to vary within the Rayleigh resolution limit.  We show the final fit to the K2 light curve in the top right panel of Fig.\,\ref{fig:hd107000} for HD\,107000 and provide all 13 determined rotation periods in Table\,\ref{tab:rotation}.  We propagated the Rayleigh limit as a conservative uncertainty for each $P_{\mathrm{rot}}$.

Several of the stars in our sample have previously been monitored to investigate their periodic variability.  We compared the literature values for the rotation period with the values derived in this work (see Table\,\ref{tab:rotation}). Most agreed well with only two exceptions.  For HD\,107000, we recovered twice the literature value.  This may occur if \citet{2012MNRAS.420..757W} performed a frequency extraction instead of fitting a high-order sine model to the \textsc{stereo} light curve in the time domain.  Our value for the rotation period of HD\,107000 offers a better fit to the K2 light curve.  The other exception is HD\,177765, for which \citet{2017A+A...601A..14M} indicated a rotation period longer than 5\,years due to the lack of variability in the mean magnetic field modulus.  The variability in the K2 light curve has a small amplitude and does not contain two full cycles of the derived rotation period (56\,days). We cannot exclude that this photometric variability has a different (possibly instrumental) origin other than rotation.

\begin{table}
\caption{Rotation periods from photometric modulation.  Values determined in this work were calculated from the K2 photometry. $^{a}$ \citet{2006SASS...25...47W},  $^{b}$ \citet{2012MNRAS.420..757W}, $^{c}$ \citet{2015A+A...581A.138B}, $^{d}$ \citet{2016AJ....152..104H}, $^{e}$ \citet{2017A+A...601A..14M}, $^{*}$ Length of the K2 light curve  is smaller than twice the rotation period.}
\centering
\tabcolsep=6pt
\begin{tabular}{lccl}
\hline
\hline
Star		& $P_{\mathrm{rot}}$ [d] 	&  $P_{\mathrm{rot}}$ [d]  	& Note 		\\
		& This work					&  Literature 				&		 	\\
\hline
HD\,97859	&$	0.792	\pm	0.008	$&$	0.7921	\pm	0.0002	$&$^{	b	}$\\
HD\,107000	&$	5.6	\pm	0.7	$&$	2.8187	\pm	0.0002	$&$^{	b	}		$\\
HD\,134759	&$				$&$	3.099	\pm	0.001	$&$^{	b	}		$\\
HD\,139160	&$				$&$	6.62	\pm	0.02	$&$^{	b	}				$\\
HD\,152366	&$	3.23	\pm	0.01	$&$				$&$^{		}				$\\
HD\,152834	&$	4.4	\pm	0.3	$&$	4.2928	\pm	0.0003	$&$^{	d	}		$\\
HD\,155127	&$	5.5	\pm	0.4	$&$	5.5243	\pm	0.0004	$&$^{	d	}		$\\
HD\,158596	&$	2.02	\pm	0.02	$&$	2.02206	\pm	0.00005	$&$^{	c	}		$\\
HD\,164224	&$	0.73	\pm	0.01	$&$				$&$^{		}				$\\
HD\,165972	&$	2.8	\pm	0.1	$&$	2.7596	\pm	0.0001	$&$^{	d	}		$\\
HD\,166804	&$	3.7	\pm	0.2	$&$	3.7035	\pm	0.0002	$&$^{	d	}		$\\
HD\,173406	&$	4.6	\pm	0.3	$&$	5.095			$&$^{	a	}			$\\
HD\,173657	&$	1.94	\pm	0.05	$&$	1.93789	\pm	0.00005	$&$^{	d	}		$\\
HD\,177013	&$	4.9	\pm	0.3	$&$	4.873			$&$^{	a	}			$\\
HD\,177562	&$				$&$				$&$^{		}				$\\
HD\,177765	&$	56	\pm	40	$&$	\gg 5$\,years&$^{	e, *	}				$\\
\hline
\end{tabular}
\label{tab:rotation}
\end{table}

\subsection{Remaining periodic variability}

By removing the rotation signal from the K2 photometry, we drastically reduced the power in the Lomb-Scargle periodograms (e.g., the bottom left panel of Fig.\,\ref{fig:hd97859}).  The strongest remaining signal within the periodogram of the residuals falls within one of the following categories:

\begin{itemize}
\item Low-frequency power excess related to imperfectly reduced and corrected K2 photometry.  This might have been caused by stitching different K2 (sub)-campaigns together into one light curve or because of an incomplete removal of long-term instrumental variability. HD\,152834 and HD\,165792 show examples of this power excess in the periodogram of the residuals.

\item Remaining aliasing structure around the ${\sim}$6\,h K2 thruster fire frequency (or its harmonics) or at the harmonics of the ${\sim}$6\,h thruster fire frequency itself.  A clear example of the former was observed for HD\,173657.

\item Remaining structure around the rotation frequency or its frequency harmonics.  This is observed, for example, for HD\,155127, HD\,158596, or HD\,166804.  The surface abundance inhomogeneities are anticipated to be stable over the time scale of the light curves, as they are linked to the stable large-scale magnetic fields.  However, spot migration could have occurred, but is considered to be unlikely.
\end{itemize}

Three stars are exceptions to these results with significant power in their residual periodograms.  Both HD\,158596 and HD\,177765 show a clear, isolated frequency peak at $17.001 \pm 0.005$\,d$^{-1}$ and $11.76 \pm 0.01$\,d$^{-1}$, respectively.  These could be Nyquist aliases of unresolved, high-frequency pulsation modes in the rare subgroup of roAp stars \citep{1982MNRAS.200..807K}.  \citet{2012MNRAS.421L..82A} discussed the detection of a 61.01\,d$^{-1}$ roAp pulsation in the spectroscopy of HD\,177765, supporting the hypothesis that we may indeed be dealing with an alias in the photometry.  We cannot exclude that the frequency peak for HD\,158596 originates in the fainter secondary component.  Finally, we note that HD\,164224 reveals a low-frequency power excess, which could be due to unresolved g-mode pulsation frequencies or internal gravity waves, which are predicted in stars with convective cores \citep[e.g.,][]{2015ApJ...806L..33A}.

We did not find evidence of frequency peaks related to stellar oscillations driven by the $\kappa$-mechanism for any of the 13 targets. 

\section{Estimation of stellar parameters}
\label{sec:spectroscopy}

To perform the magnetometric analysis, we required estimates for the effective
temperature, $T_{\mathrm{eff}}$, and the surface gravity, $\log g$, to construct
average spectra with the Least-Squares Deconvolution (LSD) technique
\citep{1997MNRAS.291..658D}.  This method drastically increased the S/N of a
possible Zeeman signature in the (LSD) Stokes\,V profile.

As a first step, we deduced the projected rotational velocity $v\sin i$ and the radial velocity for each star by fitting a rotation profile \citep[see e.g.,][]{2005oasp.book.....G} convolved with a Gaussian profile to an unblended \textsc{Fe} or \textsc{Si} line.  The Gaussian profile captures the non-rotational broadening $v_{\rm NR}$ acting upon the absorption line, due to macro-turbulence, micro-turbulence, waves, etc.  We followed the goodness-of-fit approach by \citet{2014A+A...562A.135S} to derive the values and confidence intervals for $v\sin i$ and $v_{\rm NR}$ and report the results in Table\,\ref{tab:sample}, keeping in mind that these two parameters are correlated.

Next, we determined $T_{\mathrm{eff}}$ and $\log g$ for each star by fitting pre-computed \textsc{syntv} local thermodynamical equilibrium synthetic spectra \citep{1996ASPC..108..198T} for \textsc{llmodels} atmosphere models \citep{2004A+A...428..993S} using the Grid Search in Stellar Parameters \citep[\textsc{gssp},][]{2015A+A...581A.129T} software to the observations.  Because of the significant chemical peculiarities for these stars, we performed the fitting simultaneously to the H$\beta$ and H$\gamma$ lines, while keeping the overall metallicity fixed at [M/H]\,=\,0.0 and the micro-turbulent velocity at $v_{\rm micro} = 2.0$\,km\,s$^{-1}$.   Moreover, both the line broadening and the radial velocity were fixed at the values in Table\,\ref{tab:sample}.  The grid of stellar parameters for \textsc{gssp} spanned from $T_{\mathrm{eff}}$ = 6000\,K up to 15000\,K, with steps of 500\,K, and $\log g$ from 3.0\,dex up to 4.5\,dex, with a step of 0.1\,dex.  The determined values for $T_{\mathrm{eff}}$ and $\log g$, as well as their statistical uncertainties, are also listed in Table\,\ref{tab:sample}.  These errors on $T_{\mathrm{eff}}$ and $\log g$ should be considered as a lower limit of the confidence intervals, because Ap stars are known to have a core-wing anomaly for their hydrogen lines \citep[see e.g.,][]{2001A+A...367..939C, 2002ApJ...578L..75K}.  For the stars with unresolved binary components, the values given may have contributions from both stars.  However, the difference in $T_{\mathrm{eff}}$ as a result of this would not significantly affect our conclusions.

Although the secondary components of HD\,134759, HD\,139160, and HD\,158596 should be well within the fiber area of ESPaDOnS, we did not find a clear indication of the presence of the secondary in the spectra. Therefore, we also fitted their respective observations with a single-star synthetic spectrum to estimate the stellar parameters.

\section{Magnetometry}
\label{sec:magnetic}

\begin{table*}
\caption{Magnetometric properties of the sample.  We provide the HJD at the middle of the spectropolarimetric sequence, as well as the total exposure time.  Furthermore, we indicate the $T_{\mathrm{eff}}$ and $\log g$ of the \textsc{vald3} line mask, the S/N in the LSD Stokes\,I profile, the Land\'{e} factor $g$, the mean wavelength $\lambda$, the integration range around the line centroid used for the longitudinal field calculation and FAP analysis, the detection status of a Zeeman signature (DD for definite detection, ND for non-detection and MD for marginal detection), and the measured longitudinal magnetic field.}
\centering
\tabcolsep=6pt
\begin{tabular}{lllrrrrrrcr}
\hline
\hline
Star		& HJD		& $t_{\mathrm{exp}}$ & $T_{\mathrm{eff}}$& $\log g$ 	& S/N	&$g$	&$\lambda$ &  int. range& Det.	& $B_l$\\
		& -2450000&& \multicolumn{2}{c}{line mask}\\
		& [d]				& [s]	& [K]		& [dex]				&		& &	[nm]&	[km s$\mathrm{^{-1}}$]	&	& [G]\\
\hline
HD97859		&	2457881.78364	&$4 \times 	1272	$&	14000	&	4.0	&	4829	&	1.192	&	519.13	&$\pm	150	$&	DD	&$	570	\pm	122	$\\
HD107000		&	2457475.92214	&$4 \times 	338	$&	7750	&	3.5		&	2836	&	1.210	&	519.94	&$\pm	50	$&	DD	&$	250	\pm	10	$\\
			&	2457497.87935	&$4 \times 	338	$&		&			&	3047	&	1.210	&	518.95	&$			$&	DD	&$	162	\pm	12	$\\
HD134759		&	2457799.17433	&$4 \times 	37	$&	12000	&	4.0	&	4941	&	1.183	&	520.89	&$\pm	70	$&	DD	&$	305	\pm	30	$\\
			&	2457801.17550	&$4 \times 	37	$&		&			&	5056	&	1.183	&	521.12	&$			$&	DD	&$	364	\pm	31	$\\
HD139160		&	2457820.00404	&$8 \times 	22	$&	13000	&	4.0	&	3133	&	1.177	&	520.43	&$\pm	45	$&	ND	&$	42	\pm	50	$\\
primary		&				&		&		&		&		&		&&$\pm	25	$&	ND	&$	-2	\pm	30	$\\
secondary	&				&		&		&		&		&		&&$\pm	35	$&	ND	&$	49	\pm	116	$\\
HD152366		&	2457818.06147	&$4 \times 	599	$&	10000	&	3.5	&	2843	&	1.196	&	518.52	&$\pm	40	$&	DD	&$	-82	\pm	14	$\\
HD152834		&	2457818.08468	&$4 \times 	267	$&	10000	&	3.5	&	1769	&	1.188	&	511.72	&$\pm	25	$&	DD	&$	228	\pm	20	$\\
HD155127		&	2457818.10943	&$4 \times 	676	$&	8000		&	3.5	&	1812	&	1.201	&	511.06	&$\pm	40	$&	DD	&$	-435	\pm	7	$\\
HD158596		&	2457818.15110	&$4 \times 	1060	$&	11000	&	4.0	&	2744	&	1.182	&	525.24	&$\pm	110	$&	DD	&$	610	\pm	46	$\\
HD164224		&	2457817.15008	&$4 \times 	1089	$&	8750		&	3.5	&	1373	&	1.197	&	515.58	&$\pm	50	$&	DD	&$	580	\pm	27	$\\
HD165972		&	2457905.94729	&$4 \times 	517	$&	13000	&	4.0	&	3144	&	1.185	&	518.54	&$\pm	60	$&	DD	&$	-326	\pm	50	$\\
HD166804		&	2457883.11294	&$4 \times 	443	$&	13000	&	4.0	&	3095	&	1.180	&	517.88	&$\pm	55	$&	DD	&$	-476	\pm	50	$\\
			&	2457886.92179	&$4 \times 	443	$&			&		&	3066	&	1.180	&	518.47	&$			$&	DD	&$	-520	\pm	50	$\\
			&	2457890.13132	&$4 \times 	443	$&			&		&	2468	&	1.180	&	519.00	&$			$&	DD	&$	-429	\pm	67	$\\
HD173406		&	2457500.13798	&$4 \times 	214	$&	13000	&	4.0	&	1882	&	1.174	&	519.73	&$\pm	70	$&	ND	&$	-44	\pm	45	$\\
HD173657		&	2457554.98727	&$4 \times 	226	$&	11000	&	4.0	&	3631	&	1.164	&	508.96	&$\pm	120	$&	ND	&$	-79	\pm	84	$\\
HD177013		&	2457905.96915	&$4 \times 	310	$&	12000	&	4.5	&	963	&	1.163	&	511.23	&$\pm	45	$&	MD	&$	217	\pm	59	$\\
HD177562		&	2457553.02367	&$4 \times 	262	$&	13000	&	3.5	&	4572	&	1.176	&	524.93	&$\pm	30	$&	ND	&$	16	\pm	18	$\\
HD177765		&	2457554.99870	&$4 \times 	120	$&	8500		&	4.5	&	1001	&	1.204	&	513.89	&$\pm	25	$&	DD	&$	1067	\pm	21	$\\
\hline
\end{tabular}
\label{tab:magnetometry}
\end{table*}

To boost the signal-to-noise ratio (S/N) of the possible Zeeman signatures in the Stokes\,V spectrum, we constructed an average line profile using the LSD technique.  A precomputed \textsc{vald3} line mask \citep{2015PhyS...90e4005R} with parameters close to the estimated stellar parameters was used for each star, and is listed in Table\,\ref{tab:magnetometry}.  All hydrogen lines and all spectral lines blended with hydrogen lines, telluric features, and known diffuse interstellar bands were removed from the line masks.  Furthermore, the depths of the lines in the masks were adjusted to match the observations \citep[a process so-called ``tweaking'', see e.g.,][]{2017MNRAS.465.2432G}.  The resulting LSD profiles are shown in the bottom right panel of Fig.\,\ref{fig:hd107000} for HD\,107000 and in Appendix\,\ref{sec:appendix} for the other stars.

Most of the LSD Stokes\,I profiles show some sort of distortion.  We attributed this profile variability to the surface abundance inhomogeneities, because the corresponding K2 photometry displays rotational modulation.

\subsection{Zeeman signature}
The False Alarm probability \citep[FAP;][]{1992A+A...265..669D,   1997MNRAS.291..658D} was utilized to determine the presence of a Zeeman signature in the LSD Stokes\,V profile.  Non-detections (NDs) have been assigned for a FAP\,$ > 10^{-1}\,\%$, definite detections (DDs) for FAP\,$ < 10^{-3}\,\%$, and marginal detections correspond to $10^{-3}\,\% <$\,FAP\,$ < 10^{-1}\,\%$.  Using these criteria, eleven stars have a DD, while four have a ND and one a MD.  We mark the detection status in Table\,\ref{tab:magnetometry}.

For each spectropolarimetric observation, we also determined the longitudinal magnetic field \citep{1979A+A....74....1R} as:
\begin{equation}
B_l = -2.14 \cdot 10^{11} \frac{\int v V(v) \mathrm{d}v}{\lambda g c \int[1-I(v)] \mathrm{d}v} \, \mathrm{,}
\label{eq:longfield}
\end{equation}
\noindent where $V(v)$ and $I(v)$ are the LSD Stokes\,V and I profiles, respectively, for a given velocity $v$ in km s$\mathrm{^{-1}}$. The mean Land\'{e} factor, $g$, and the mean wavelength (in nm), $\lambda$, result from the LSD technique.  The speed of light, $c$, is given in km s$\mathrm{^{-1}}$ and the longitudinal magnetic field, $B_l$, in Gauss.  The integration range spans the full width of the Zeeman signature and the mean absorption line profile.  We report the values for $\lambda$, $g$, the integration range, and the calculated longitudinal magnetic field in Table\,\ref{tab:magnetometry}.

The determined $B_l$ values are well in line with the expectations for magnetic hot stars with a polar strength of at least several 100\,G \citep[where the polar strength is at least 3.5 times the measured longitudinal magnetic field;][]{1967ApJ...150..547P}.  Stars for which we obtained a ND also have a measured $B_l$ compatible with zero within the 2$\sigma$ confidence interval. In addition, Zeeman splitting is observed in individual lines (Stokes\,I) of HD\,177765 \citep[as already clearly indicated in Figure\,4 of][]{1997A+AS..123..353M}, confirming that the star hosts a strong field of several kG.  This strong magnetic field agrees with the long rotation period for the star (out of the studied sample) when considering magnetic braking.

Nearly all of the detected Zeeman profiles have a typical appearance for large-scale magnetic fields with a simple geometry (dominantly dipolar), where the exact shape depends on the rotation phase and on the unknown inclination and obliquity angles.  Distortions of the Zeeman signature due to the distorted LSD Stokes\,I profile did occur for several stars, such as HD\,166804.  However, the signature was sufficiently clear that these distortions did not cause any issues with the detection.  Moreover, the rotation periods are sufficiently long compared to the total exposure time of the spectropolarimetric sequence, so that the measured $B_l$ values were not affected by these distortions.

\subsection{Binary systems} 
\label{sec:magnetic_binaries}
As mentioned earlier, three stars of our sample are part of binary systems.  A component of two of these (i.e., HD\,134759 and HD\,158596) clearly host a large-scale magnetic field. The spectropolarimetric sequence of HD\,139160, however, led to a ND.  Here, we consider several explanations.

\begin{itemize}
\item  The ND could be caused by the binary nature of HD\,139160.  The distortions in the LSD Stokes\,I profile could be due to the presence of the nearby companion, classifying HD\,139160 as SB2 system, instead of the previously claimed surface abundance inhomogeneities.  By means of LS fitting, we described the LSD Stokes\,I profile with two Gaussian functions to represent the individual components (see Fig.\,\ref{fig:hd139160}).  After subtracting either one of the Gaussian profiles, we isolated the individual components in the Stokes\,I profile.  We then used these two profiles to redetermine the FAP for a Zeeman signature (in the LSD Stokes\,V) and recomputed the longitudinal magnetic field (see Table\,\ref{tab:magnetometry}).  Again, we did not detect a signature of a large-scale magnetic field for the individual components.  Hence, the binary nature is probably not the cause of the ND.

\item HD\,139160 has a rather short orbital period of 5.28\,d, according to the modelling efforts of \citet{1987ApJS...64..487L}  to the measured radial velocities.  As magnetic hot stars seem to be less common in binary systems with short orbital periods \citep[as inferred from the BinaMIcS results,][]{2015IAUS..307..330A}, this could explain the ND.  Yet, spectropolarimetric data with a much higher S/N are needed to exclude the possibility of a weak magnetic field.

\item As indicated earlier, two sub-groups of He-weak stars exist.  The He-weak stars related to the HgMn stars do not host a large-scale magnetic field.  We compared the Stokes\,I spectrum with the synthetic \textsc{syntv} + \textsc{llmodels} (with solar abundances) spectrum, but did not find clear evidence of strong P or Ga absorption lines, which would associate HD\,139160 to this non-magnetic sub-group of He-weak stars.  We did, however, note the substantially weaker He\,\textsc{I} lines.  Thus, HD\,139160 may host a large-scale magnetic field, but that is weaker than our detection threshold.

\end{itemize}

\section{Discussion}
\label{sec:discussion}
\subsection{Rotation period}
\begin{figure}
\centering
\includegraphics[width=0.45\textwidth, height = 0.33\textheight]{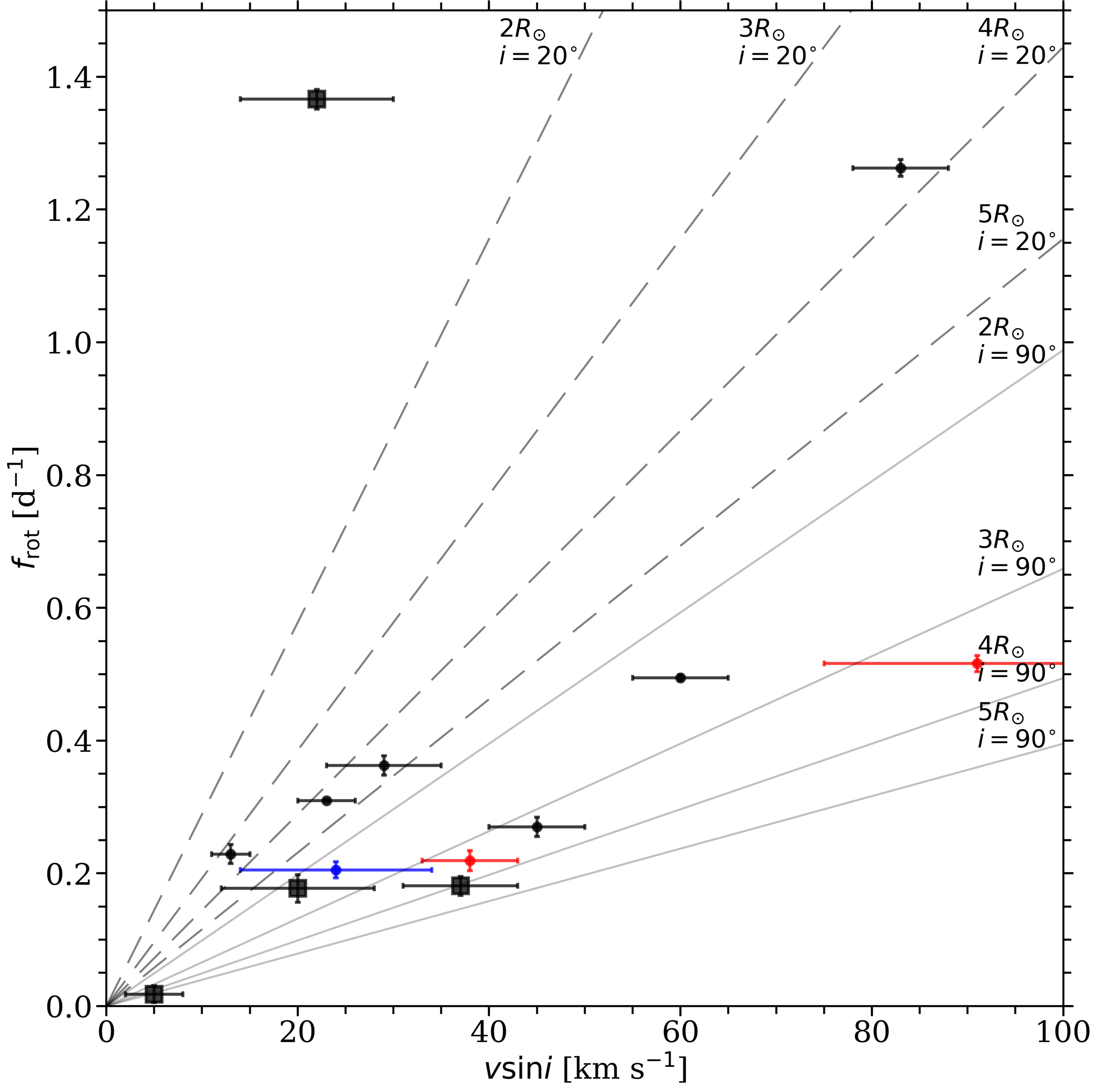}%
\caption{Rotation frequency $f_{\mathrm{rot}}$ versus $v\sin i$ (dots).  The colour of the symbols indicates the magnetic detection (black for a DD, blue for a MD, red for a ND).  The symbol relates to the measured $T_{\mathrm{eff}}$, with squares for A-type stars and dots for B-type stars. Solid lines represent the anticipated rotation frequency for the labelled stellar radius, inclination angle, and measured $v\sin i$.  Solid lines are for $i=90\,^{\circ}$ and dashed lines for $i=20\,^{\circ}$.  Anticipated rotation frequencies at a given $v\sin i$ decrease for increasing model radii ($2R_{\odot}$, $3R_{\odot}$, $4R_{\odot}$, and $5R_{\odot}$, respectively). We indicate the confidence intervals for both $f_{\mathrm{rot}}$ and $v\sin i$, where we added the errors of $v\sin i$ and of $v_{\rm NR}$ in quadrature given the correlated nature of these two parameters.}
  \label{fig:prot_vsini}
\end{figure}

The determined rotation periods indicate that all stars are slow to moderate rotators.  This result was expected because of a known bias introduced in the sample star selection since we required to obtain high S/N, high-resolution spectropolarimetry with reasonable exposure times. Nevertheless, rotation periods of a few days are typical of Ap/Bp stars.  We compare the determined rotation frequencies with the derived values for $v\sin i$ in Fig.\,\ref{fig:prot_vsini}.  No strict relation is present, as expected due to differences in stellar radii and the anticipated random distribution of inclination angles.  HD\,97859 and HD\,164224 seem closer to (rotation) pole-on than the others, as they are situated in the upper part of the diagram.  The two probable roAp stars do not reside in a particular region of Fig.\,\ref{fig:prot_vsini}.  The variety in the inclination angle $i$, and the obliquity angle $\beta$, is further supported by the different shapes of the rotational modulation in the K2photometry.

\subsection{Magnetic detections and non-detections}

Out of the 16 expected magnetic stars in our sample, we directly detected the presence of a strong, large-scale magnetic field for 11 stars.  This is the first firm detection for 9 of these stars and the first for all 11 stars employing high-resolution spectropolarimetry.  We have a 69\,\% detection rate in our sample, with no obvious correlation with the stellar properties for the detection status.  The spectropolarimetric measurements for the four stars with a ND and the one MD are of similar quality as the other 11 stars with detected magnetic fields.  Moreover, it seems unlikely that the NDs/MDs were due to particularly unfavourable rotation phases where the field is seen in a cross-over configuration (see top right panels of Figures in Appendix\,\ref{sec:appendix}).

Two of the stars with a ND, namely HD\,173406 and HD\,173657, show stable rotational modulation over the duration of their K2 light curves.  Because of the poor quality light curve of HD\,177562, we could not confirm the presence of rotational modulation for this star with a ND.  Yet, the comparison of the ESPaDOnS spectra with their respective synthetic spectra suggests chemical peculiarities, as claimed by the \citet{2009A+A...498..961R} catalogue.  Therefore, it is likely that these three stars are indeed Ap/Bp stars with expected, stable surface abundance inhomogeneities, and thus host a large-scale magnetic field.  It may be that these stars host a weaker large-scale magnetic field than what is typically anticipated for Ap/Bp stars, which is in agreement with the increased number of detections of such weak large-scale magnetic fields \citep[e.g., HD\,5550;][]{2016A+A...589A..47A}.  These weaker magnetic fields could be an evolutionary consequence if the stars have burned a considerable amount of their hydrogen, while still undergoing hydrogen core burning, leading to an increased stellar radius and assuming magnetic flux conservation \citep[e.g.,][]{2008A+A...481..465L}.  This scenario could be further enhanced if magnetic field decay has occurred \citep[as determined by][for the most massive magnetic stars]{2016A+A...592A..84F}.

The distortions of the LSD Stokes\,I profile of HD\,173657 (see bottom right panel of Fig.\,\ref{fig:hd173657}) could be due to a resolved SB2 system, instead of surface abundance inhomogeneities.  However, this star is not known to be a binary system and we only obtained one observation.  More spectroscopic observations are needed to corroborate this hypothesis.

We also did not detect the presence of a large-scale magnetic field for the He-weak star HD\,139160.  We explored the possibility of a large-scale magnetic field in the individual components of the binary system, but again obtained a ND for both components.  Since we did not note any strong P or Ga lines in the spectroscopic observations, HD\,139160 is inferred to be unrelated to be related to the non-magnetic HgMn stars.  Thus, the presence of a large-scale magnetic field is anticipated, that is likely weaker than what was expected when computing the exposure time for the spectropolarimetric sequences.

\subsection{Lack of heat-driven stellar oscillations}

We found no evidence for the presence of stellar oscillations produced by the $\kappa$ mechanism in any of the Ap/Bp stars studied in this work. Some of our target stars do reside in the theoretical $\kappa$-driven instability regions. Although only a small fraction of the stars in these regions are known to have observable oscillations at current detection thresholds, the fact that none of our targets show such pulsations is significant.  These results provide support for the theoretical predictions that strong large-scale magnetic fields change the nature of waves and may even completely suppress them \citep{2017MNRAS.466.2181L}.

Two stars in our sample (HD\,158596 and HD\,177765) show statistically-significant peaks in their periodograms with strong evidence that there are Nyquist aliases of high-frequency roAp pulsations.  Furthermore, HD\,164224 shows a significant low-frequency power excess that is unlikely to be explained by instrumental effects and could be caused by unresolved g-mode pulsation modes or internal gravity waves.  Extended observations of all these candidate pulsators will be extremely useful in characterising the exact nature of their observed variability in the K2 photometry.

Several pulsating magnetic hot stars are known, such as $\beta$\,Cep \citep{2000ApJ...531L.143S, 2013A+A...555A..46H}, V2052\,Oph \citep{2012A+A...537A.148N, 2012MNRAS.424.2380H, 2012MNRAS.427..483B}, $\xi^1$\,CMa \citep{2017MNRAS.471.2286S}, HD\,43317 \citep{2012A+A...542A..55P, 2013A+A...557L..16B, 2017A+A...605A.104B}, HD\,188774 \citep{2015MNRAS.454L..86N}, $\beta$\,Cen\,Ab \citep{2011A+A...536L...6A, 2016A+A...588A..55P}, and $\rho$\,Pup \citep{2017MNRAS.468L..46N}. However, all these stars host a rather weak magnetic field. The only strongly magnetic star exhibiting variations interpreted as $\kappa$-driven pulsations known at present is the He-strong star HD\,96446, with a dipolar field strength of about 5\,kG \citep{2012A+A...546A..44N, 2017MNRAS.464L..85J}.

\section{Summary and conclusions}
\label{sec:conclusions}

We observed 16 chemically peculiar hot stars with space-based high-precision photometry to study their periodic variability, and with modern high-resolution optical spectropolarimetry to detect the signatures of a large-scale magnetic field.  The investigated stars are all slow to moderate rotators, with large-scale magnetic fields being easier to detect from spectropolarimetry in slowly-rotating stars compared to fast rotators. Although our sample of stars was biased towards slow rotators, the observed rotation periods are typical of Ap/Bp stars.  It is known that gravity-mode oscillations are easier to detect in slowly-rotating OB-type stars \citep{2003SSRv..105..453A, 2017ApJ...847L...7A}. Therefore, our sample bias should favour the detection of such oscillations.

In our sample, 13 stars have a high-quality K2 light curve and all of them show evidence of rotational modulation.  We detected a large-scale magnetic field for 69\% of our sample stars and found no obvious correlation between this detection and the stellar parameters, such as binarity, rotation rate, or occurrence of possible stellar pulsations.  All four non-magnetic stars, HD\,139160, HD\,173406, HD\,173657, and HD\,177562 show either rotational modulation, peculiar chemical surface abundances, or both, and therefore likely host an undetected weak large-scale magnetic field.

No $\kappa$-driven oscillations were detected in any of the targets in oursample. Two stars, namely HD\,158596 and HD\,177765, show an isolated high-frequency peak in their periodogram, which we interpret as the Nyquist alias of high-frequency roAp pulsations.  One star, HD\,164224, shows a significant low-frequency power excess in its periodogram which could be evidence of convectively-driven internal gravity waves.  Further observations and analyses of these candidate pulsators are needed to fully test these hypotheses, since we are too limited by the length and cadence of the K2 space photometry.  

We conclude that, after major efforts with the K2 mission and modern spectropolarimetry, the best stars to try and perform in-depth magneto-asteroseismic modelling are those with a moderate magnetic field which does not suppress $\kappa$-driven modes. The best target for this is and remains the CoRoT target HD\,43317, for which a series of gravity modes with consecutive radial order has been found in the CoRoT photometry \citep{2012A+A...542A..55P, 2013A+A...557L..16B} and whose magnetic field detection and modelling led to fully consistent results between the seismically and magnetically derived rotation frequency of the star \citep{2017A+A...605A.104B}.


\section*{Acknowledgements}
B.B. would like to thank Dr. Andrew Tkachenko for the useful discussions on GSSP.

This work has made use of the VALD database, operated at Uppsala University, the Institute of Astronomy RAS in Moscow, and the University of Vienna. This research has made use of the SIMBAD database operated at CDS, Strasbourg (France), and of NASA's Astrophysics Data System (ADS). Some of the data presented in this paper were obtained from the Mikulski Archive for Space Telescopes (MAST). STScI is operated by the Association of Universities for Research in Astronomy, Inc., under NASA contract NAS5-26555. Support for MAST for non-HST data is provided by the NASA Office of Space Science via grant NNX09AF08G and by other grants and contracts. This paper includes data collected by the Kepler mission. Funding for the Kepler mission is provided by the NASA Science Mission directorate. Part of the research leading to these results has received funding from the European Research Council (ERC) under the European Union's Horizon 2020 research and innovation programme (grant agreement N$^\circ$670519: MAMSIE) and from the Research Foundation Flanders (FWO, grant agreement G.0B69.13).




\bibliographystyle{mnras}
\bibliography{PhD_ADS} 



\appendix

\section{Figures}
\label{sec:appendix}
Figures, showing the K2 photometry, rotational modulation, and the LSD profiles for all stars except HD\,107000, are provided here. HD\,107000 is shown in Fig.\,\ref{fig:hd107000}.

\begin{figure}
		\centering
			\includegraphics[width=0.45\textwidth, height = 0.33\textheight]{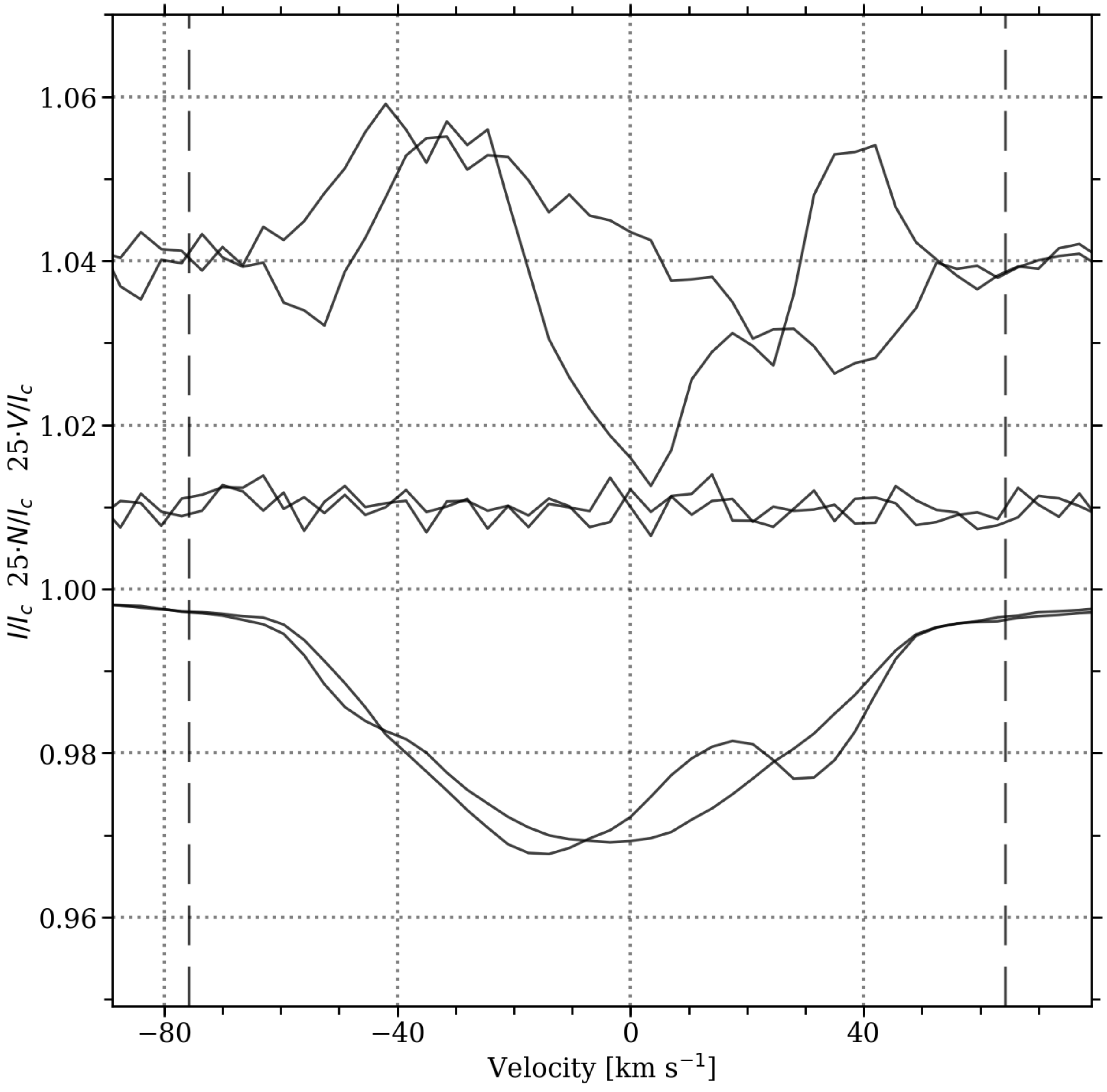}%
			\caption{LSD profiles of the ESPaDOnS spectropolarimetry for HD\,134759 where the Stokes\,V, the diagnostic null, and the Stokes\,I profiles are shown, with an offset for increased visibility. Integration limits for the determination of the FAP and longitudinal magnetic field are given by the black lines.}
			\label{fig:hd134759}
\end{figure}
\begin{figure}
		\centering
			\includegraphics[width=0.45\textwidth, height = 0.33\textheight]{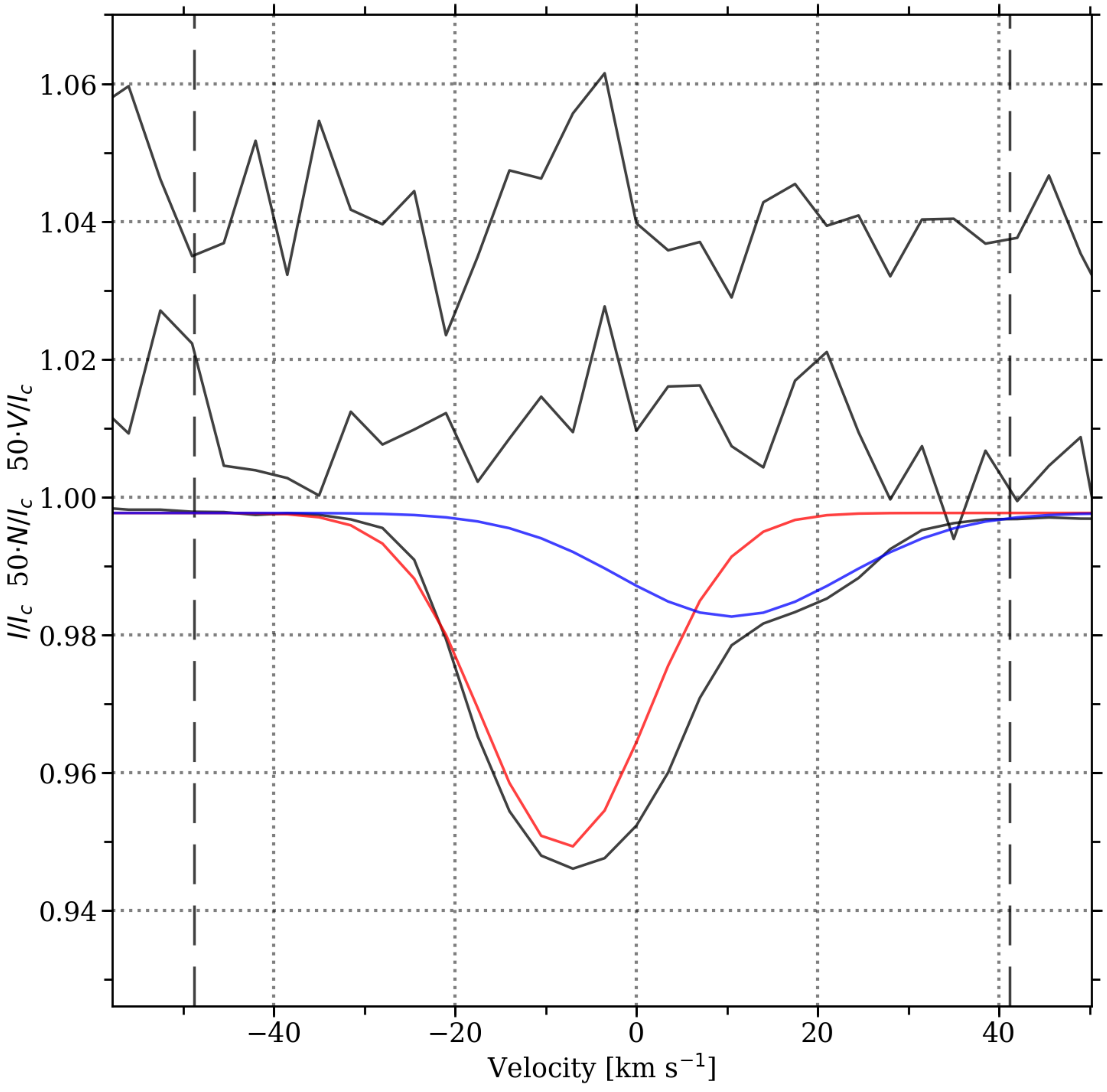}%
			\caption{LSD profiles of the ESPaDOnS spectropolarimetry for HD\,139160. No magnetic field was detected in the spectropolarimetry.  The two individual Gaussian models for the LSD Stokes\,I for the primary (red) and secondary (blue) component are also indicated.}
			\label{fig:hd139160}
\end{figure}

\begin{figure*}
		\centering
			\includegraphics[width=\textwidth, height = 0.33\textheight]{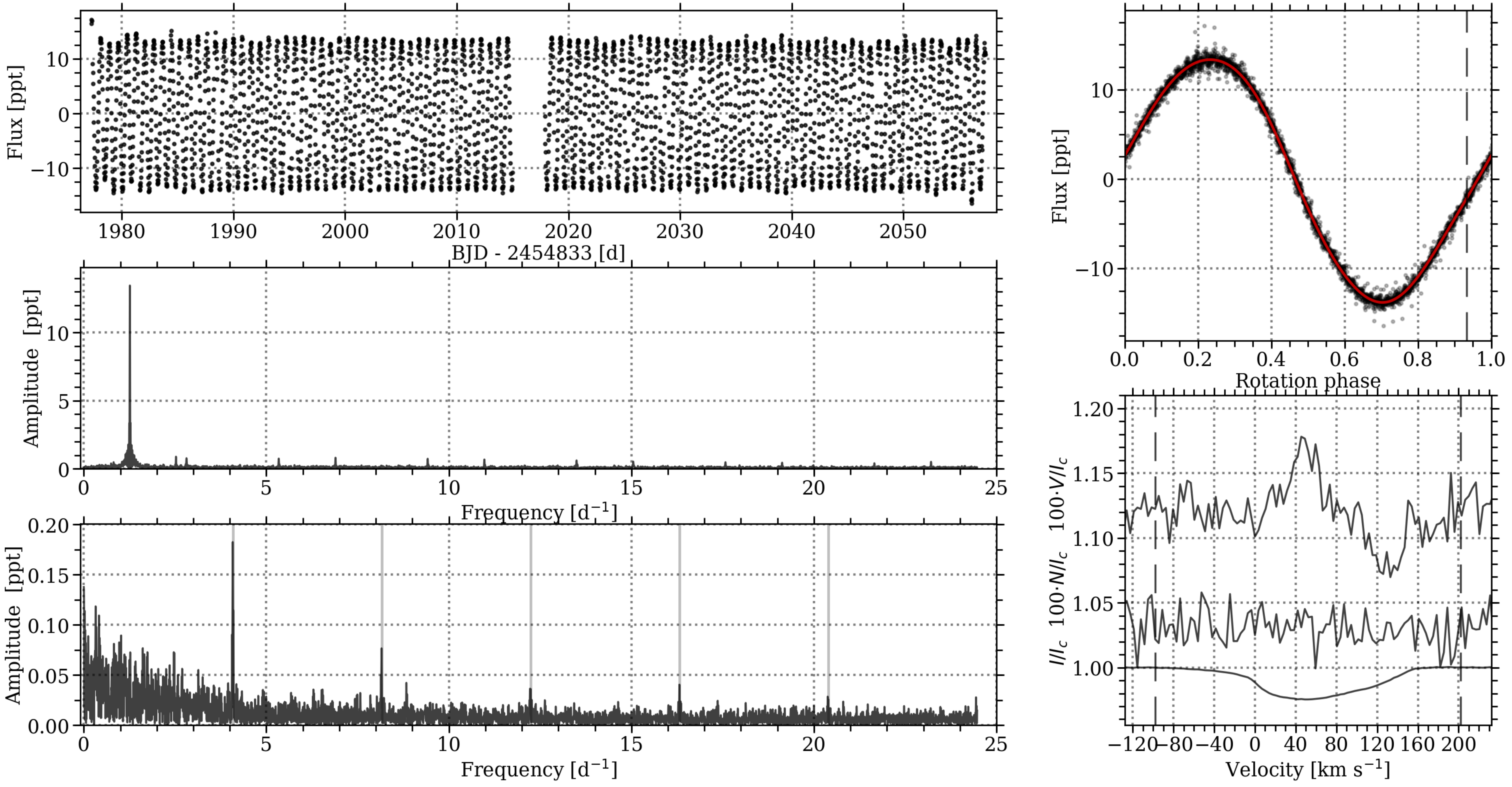}%
			\caption{Information related to HD\,97859, showing the K2 photometry, rotational modulation, and the LSD profiles of the ESPaDOnS data.  Same color coding was applied as Fig.\,\ref{fig:hd107000}.}
			\label{fig:hd97859}
\end{figure*}
\begin{figure*}
		\centering
			\includegraphics[width=\textwidth, height = 0.33\textheight]{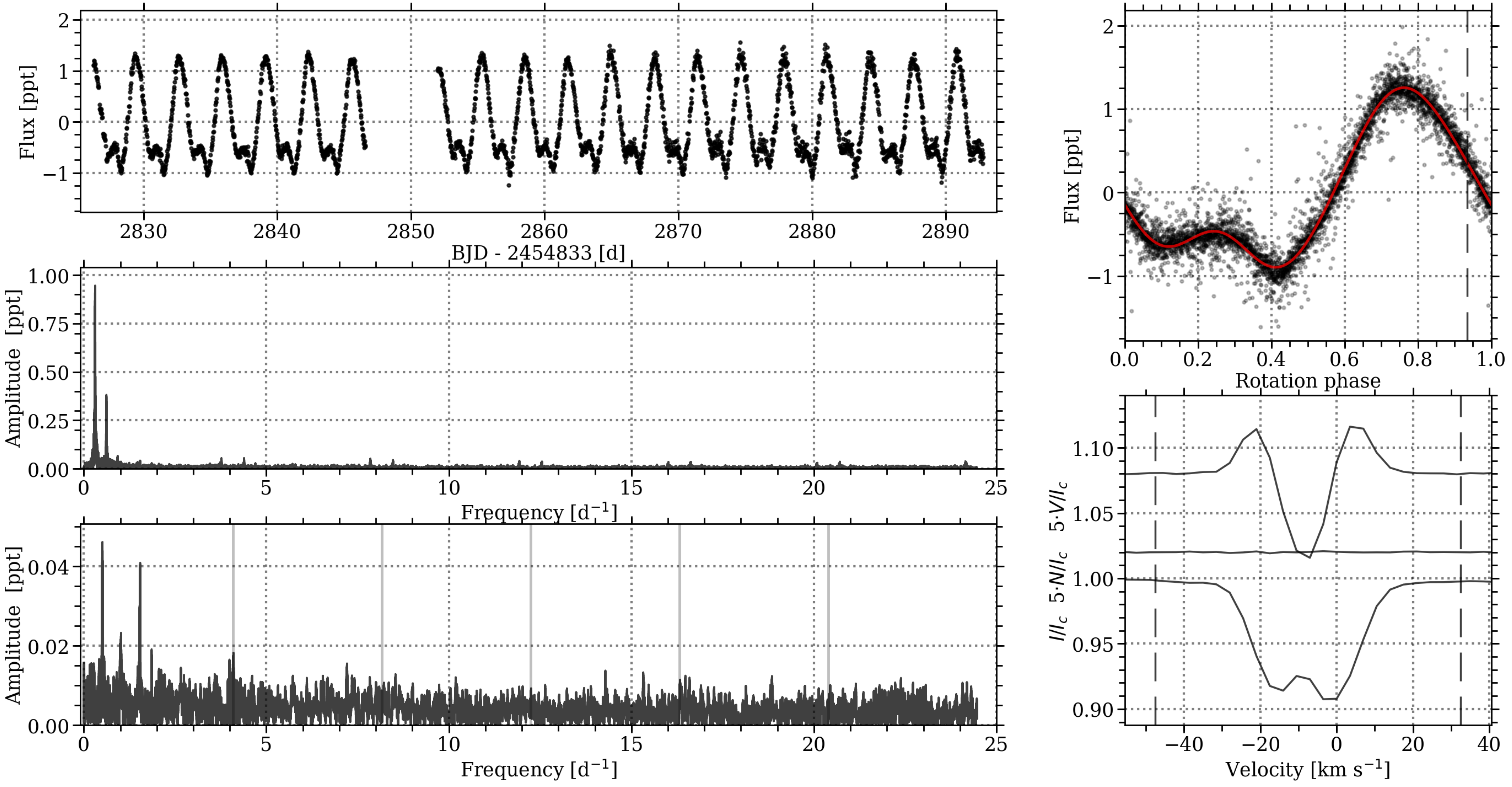}%
			\caption{Information related to HD\,152366, showing the K2 photometry, rotational modulation, and the LSD profiles of the ESPaDOnS data.  Same color coding was applied as Fig.\,\ref{fig:hd107000}.  Only the photometry of C11 was shown in the \textit{top left} panel, while the complete K2 light curve was employed for the rotational modulation, and periodograms.}
			\label{fig:hd152366}
\end{figure*}
\begin{figure*}
		\centering
			\includegraphics[width=\textwidth, height = 0.33\textheight]{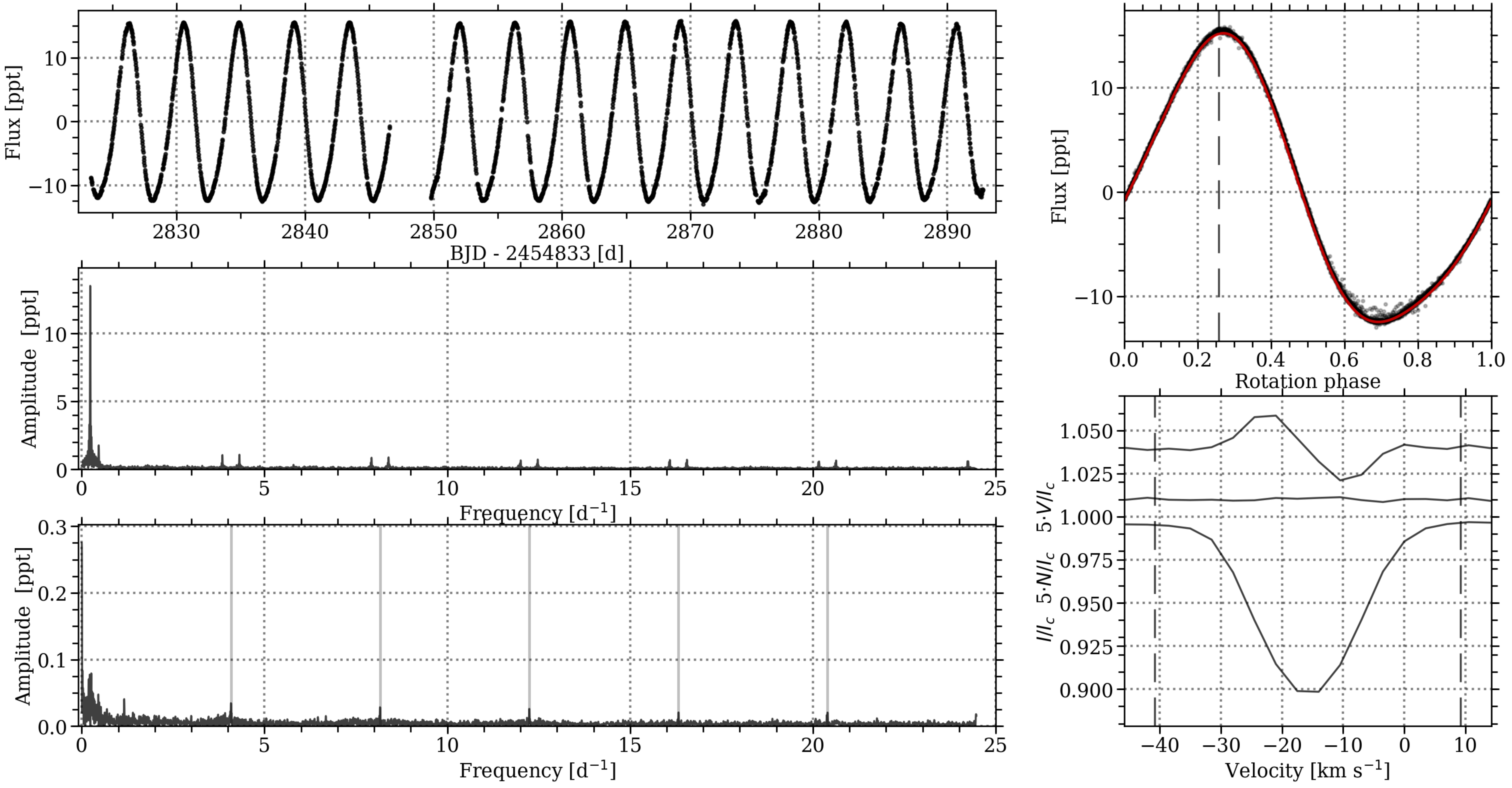}%
			\caption{Information related to HD\,152834, showing the K2 photometry, rotational modulation, and the LSD profiles of the ESPaDOnS data.  Same color coding was applied as Fig.\,\ref{fig:hd107000}.}
			\label{fig:hd152834}
\end{figure*}
\begin{figure*}
		\centering
			\includegraphics[width=\textwidth, height = 0.33\textheight]{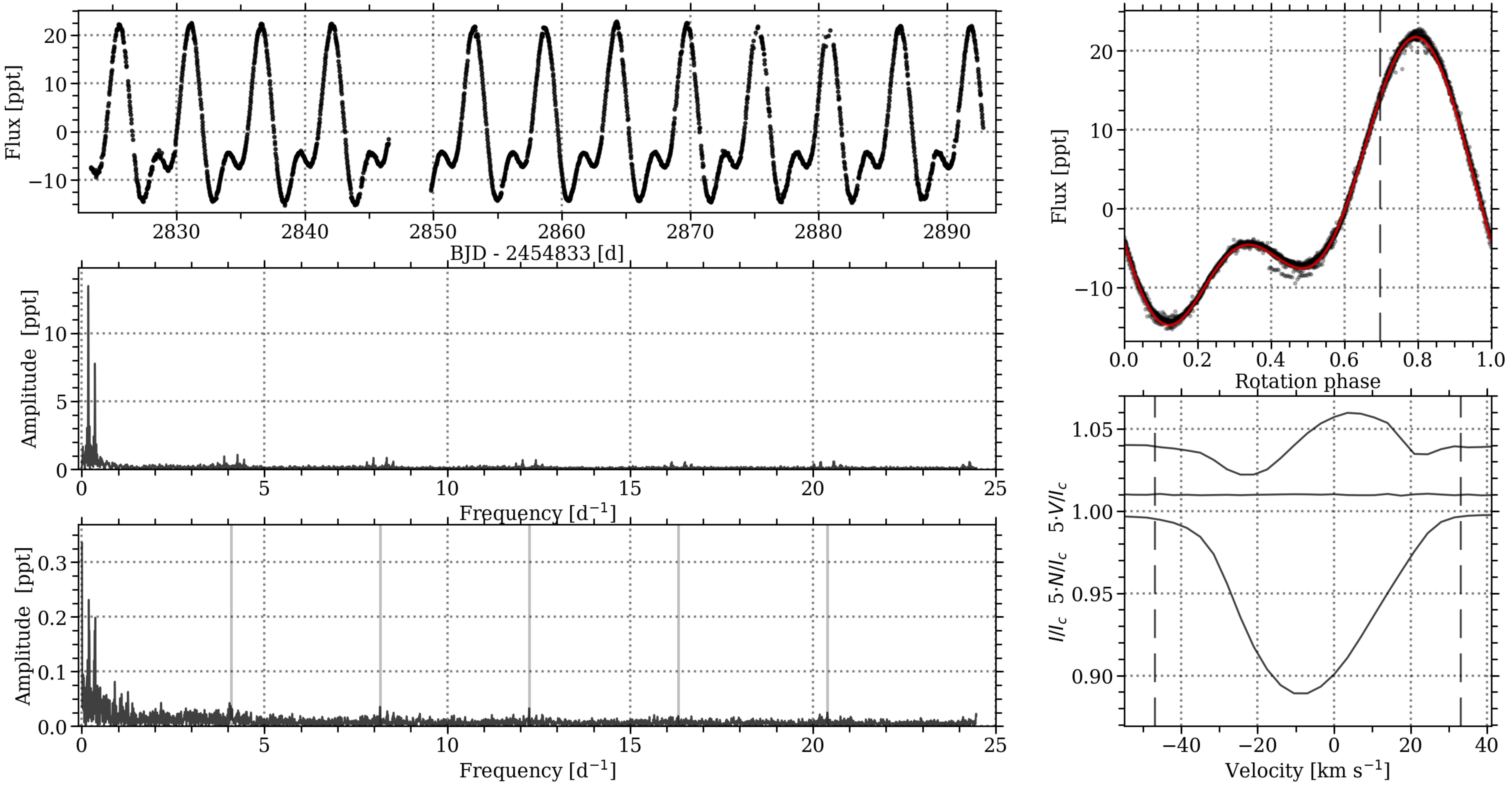}%
			\caption{Information related to HD\,155127, showing the K2 photometry, rotational modulation, and the LSD profiles of the ESPaDOnS data.  Same color coding was applied as Fig.\,\ref{fig:hd107000}.}
			\label{fig:hd155127}
\end{figure*}
\begin{figure*}
		\centering
			\includegraphics[width=\textwidth, height = 0.33\textheight]{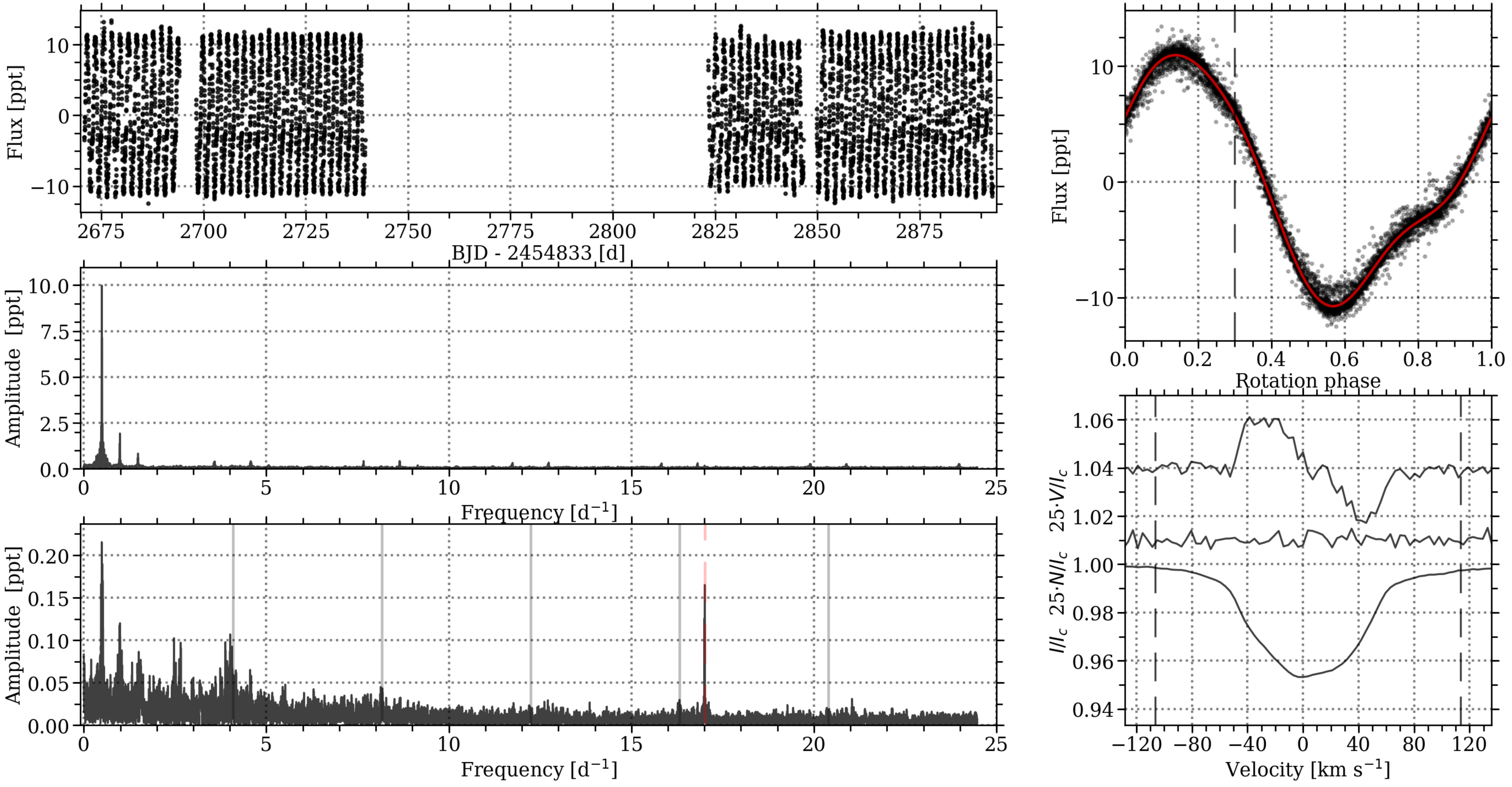}%
			\caption{Information related to HD\,158596, showing the K2 photometry, rotational modulation, and the LSD profiles of the ESPaDOnS data.  Same color coding was applied as Fig.\,\ref{fig:hd107000}.  The alias frequency of the presumed roAp pulsation was marked by the red dashed line in the \textit{bottom left} panel.}
			\label{fig:hd158597}
\end{figure*}
\begin{figure*}
		\centering
			\includegraphics[width=\textwidth, height = 0.33\textheight]{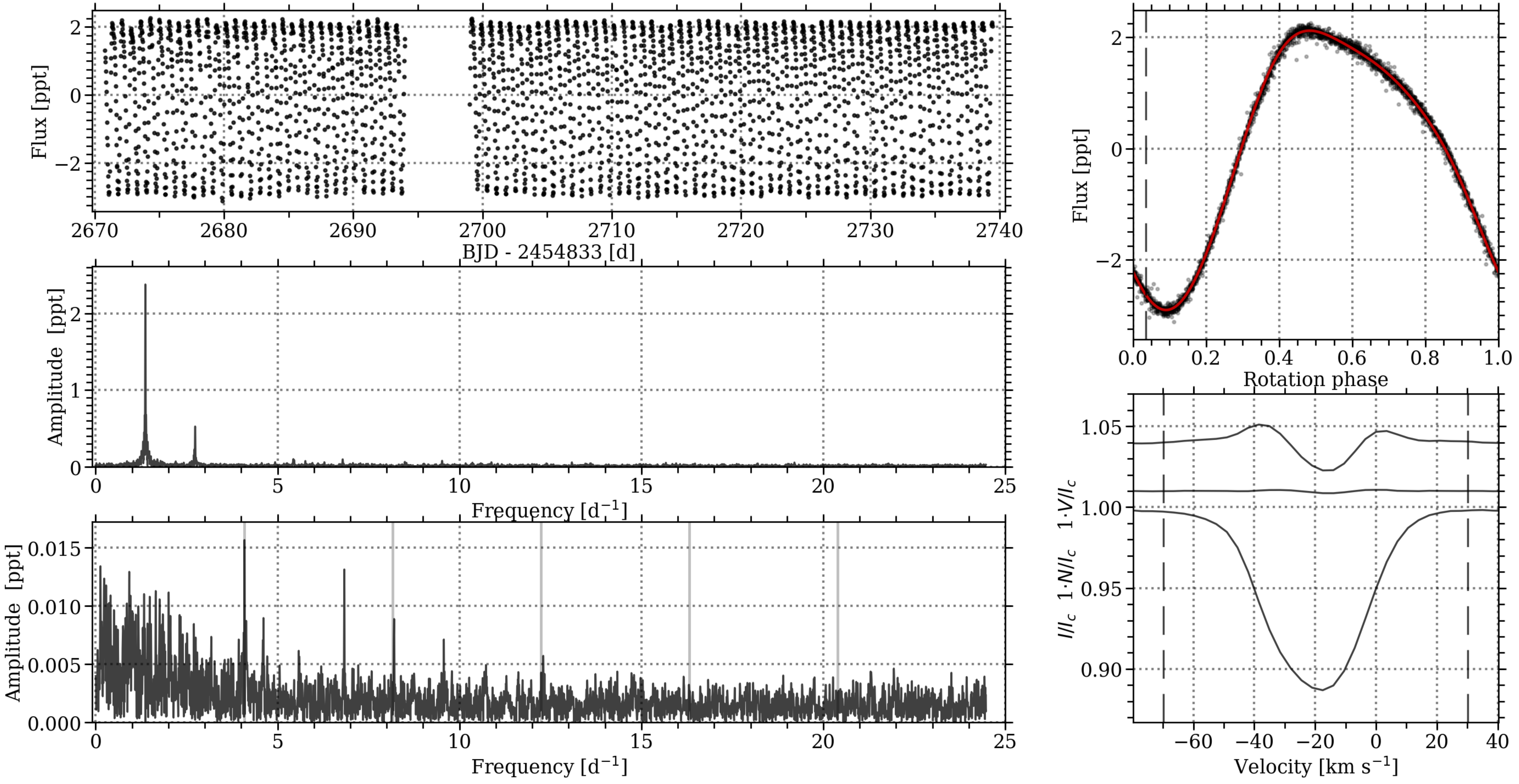}%
			\caption{Information related to HD\,164224, showing the K2 photometry, rotational modulation, and the LSD profiles of the ESPaDOnS data.  Same color coding was applied as Fig.\,\ref{fig:hd107000}.  Some addition low-frequency power excess remained in the periodogram of the K2 residuals, suggesting unresolved g mode frequencies.}
			\label{fig:hd164224}
\end{figure*}
\begin{figure*}
		\centering
			\includegraphics[width=\textwidth, height = 0.33\textheight]{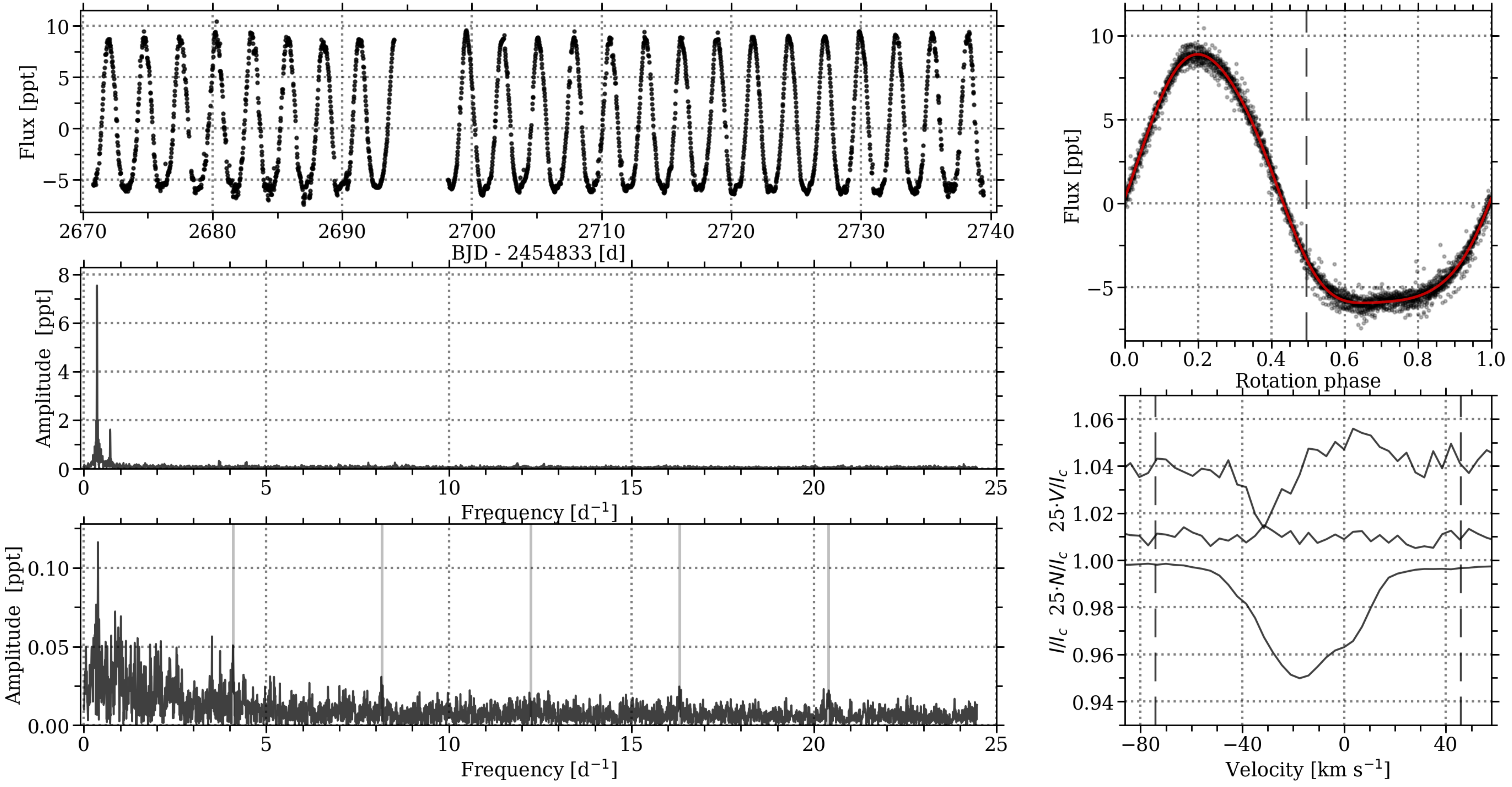}%
			\caption{Information related to HD\,165792, showing the K2 photometry, rotational modulation, and the LSD profiles of the ESPaDOnS data.  Same color coding was applied as Fig.\,\ref{fig:hd107000}.}
			\label{fig:hd165972}
\end{figure*}
\begin{figure*}
		\centering
			\includegraphics[width=\textwidth, height = 0.33\textheight]{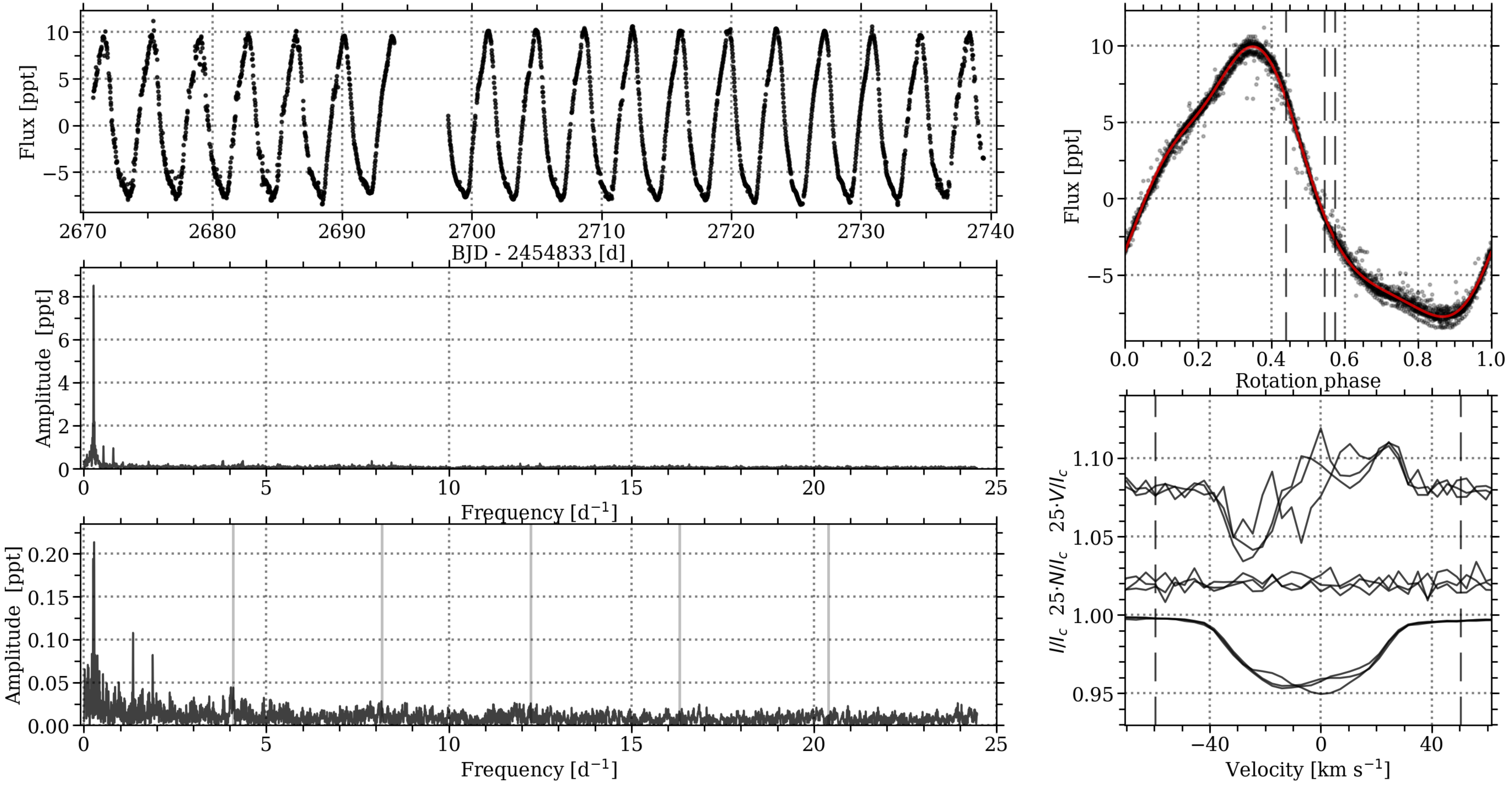}%
			\caption{Information related to HD\,166804, showing the K2 photometry, rotational modulation, and the LSD profiles of the ESPaDOnS data.  Same color coding was applied as Fig.\,\ref{fig:hd107000}.}
			\label{fig:hd166804}
\end{figure*}
\begin{figure*}
		\centering
			\includegraphics[width=\textwidth, height = 0.33\textheight]{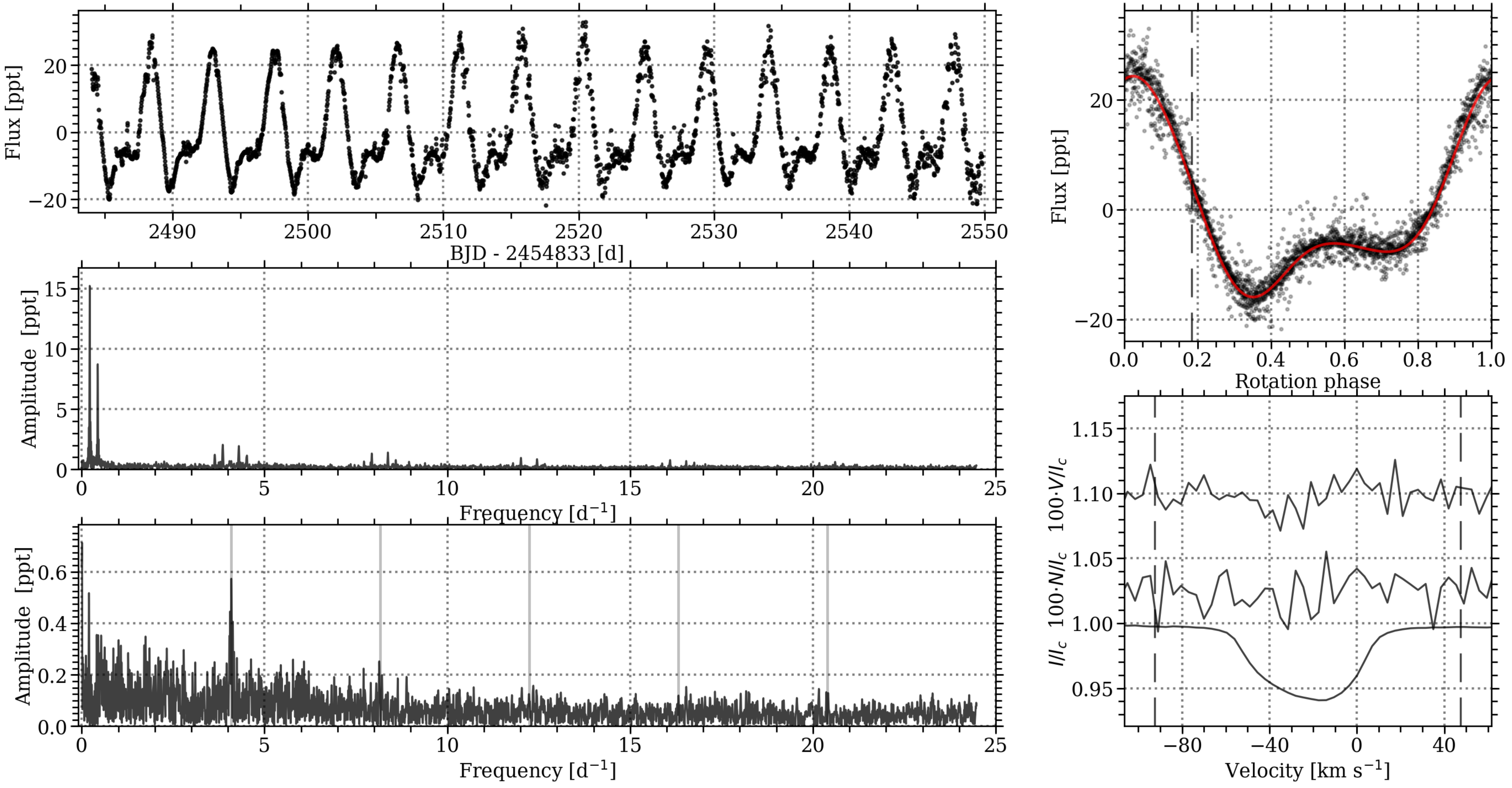}%
			\caption{Information related to HD\,173406, showing the K2 photometry, rotational modulation, and the LSD profiles of the ESPaDOnS data.  Same color coding was applied as Fig.\,\ref{fig:hd107000}.  No magnetic field was detected in the spectropolarimetry.}
			\label{fig:hd173406}
\end{figure*}
\begin{figure*}
		\centering
			\includegraphics[width=\textwidth, height = 0.33\textheight]{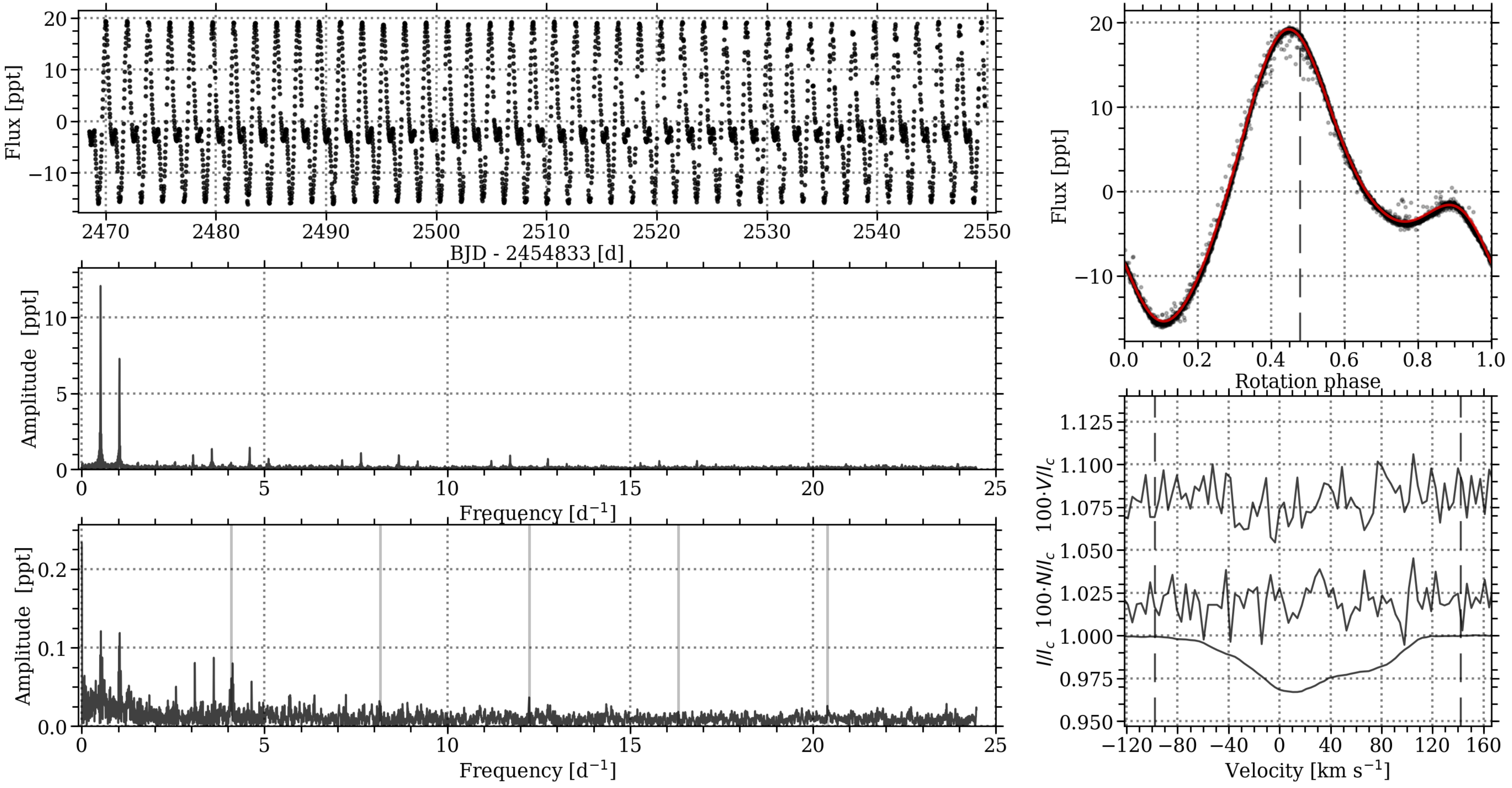}%
			\caption{Information related to HD\,173657, showing the K2 photometry, rotational modulation, and the LSD profiles of the ESPaDOnS data.  Same color coding was applied as Fig.\,\ref{fig:hd107000}.  No magnetic field was detected in the spectropolarimetry.}
			\label{fig:hd173657}
\end{figure*}
\begin{figure*}
		\centering
			\includegraphics[width=\textwidth, height = 0.33\textheight]{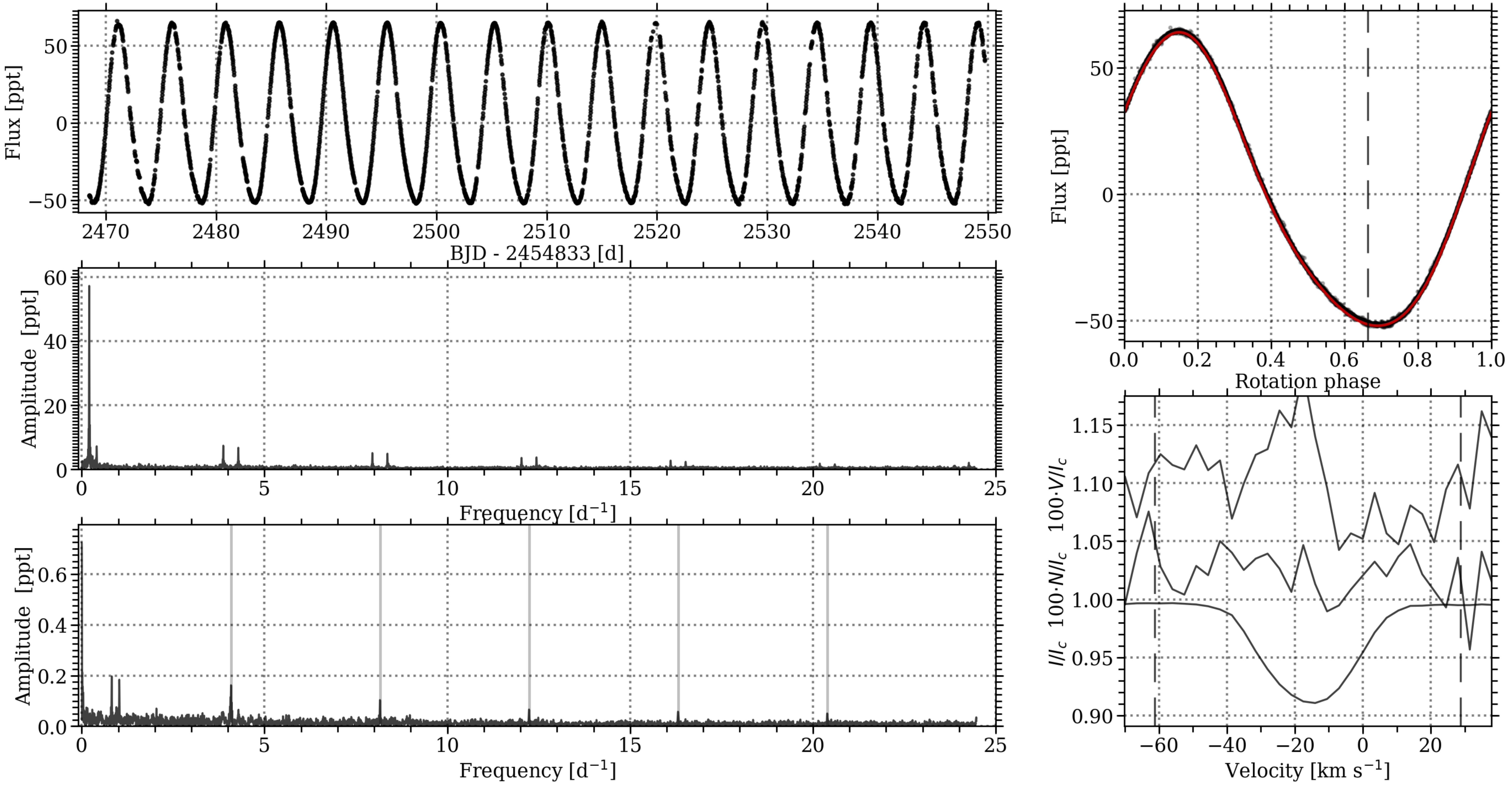}%
			\caption{Information related to HD\,177013, showing the K2 photometry, rotational modulation, and the LSD profiles of the ESPaDOnS data.  Same color coding was applied as Fig.\,\ref{fig:hd107000}.  }
			\label{fig:hd177013}
\end{figure*}
\begin{figure*}
		\centering
			\includegraphics[width=\textwidth, height = 0.25\textheight]{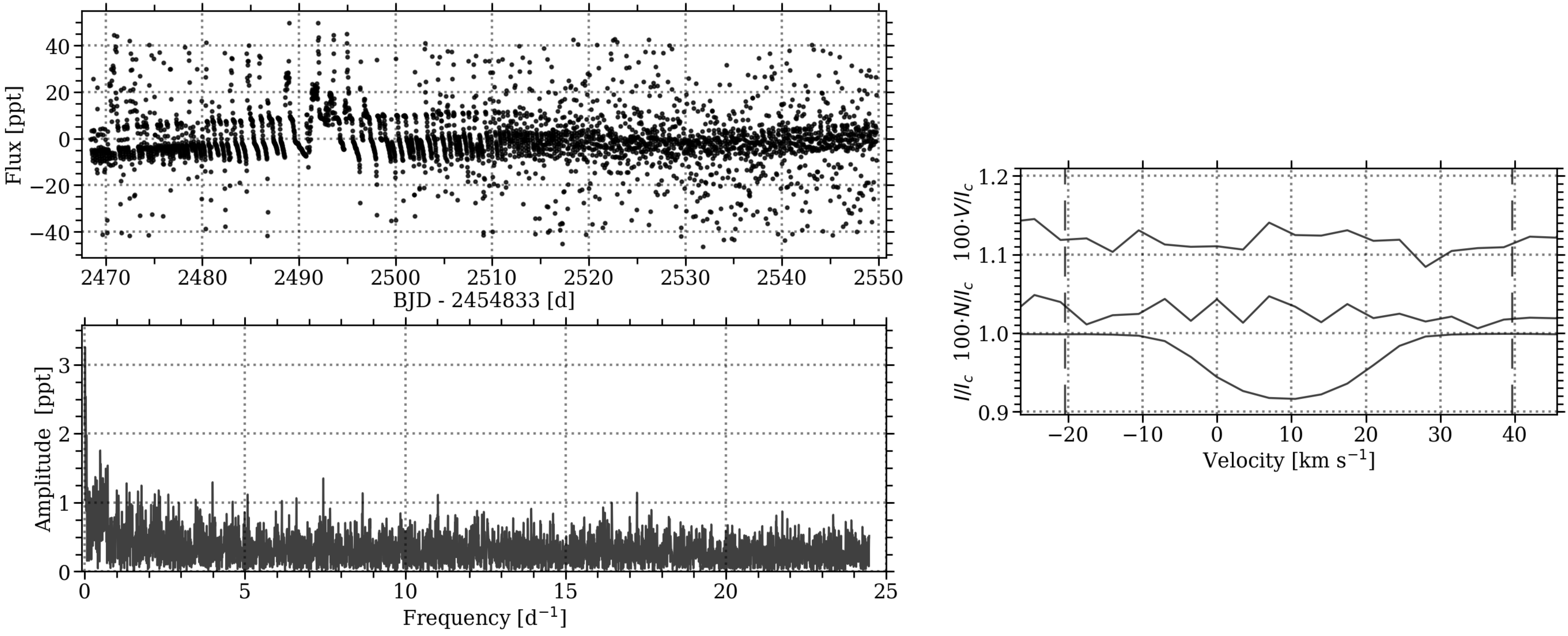}%
			\caption{Information related to HD\,177562, showing the K2 photometry, and the LSD profiles of the ESPaDOnS data.  Same color coding was applied as Fig.\,\ref{fig:hd107000}.  Since the K2 photometry was of poor quality, we were unable to recover any periodic variability. No magnetic field was detected in the spectropolarimetry.}
			\label{fig:hd177562}
\end{figure*}
\begin{figure*}
		\centering
			\includegraphics[width=\textwidth, height = 0.33\textheight]{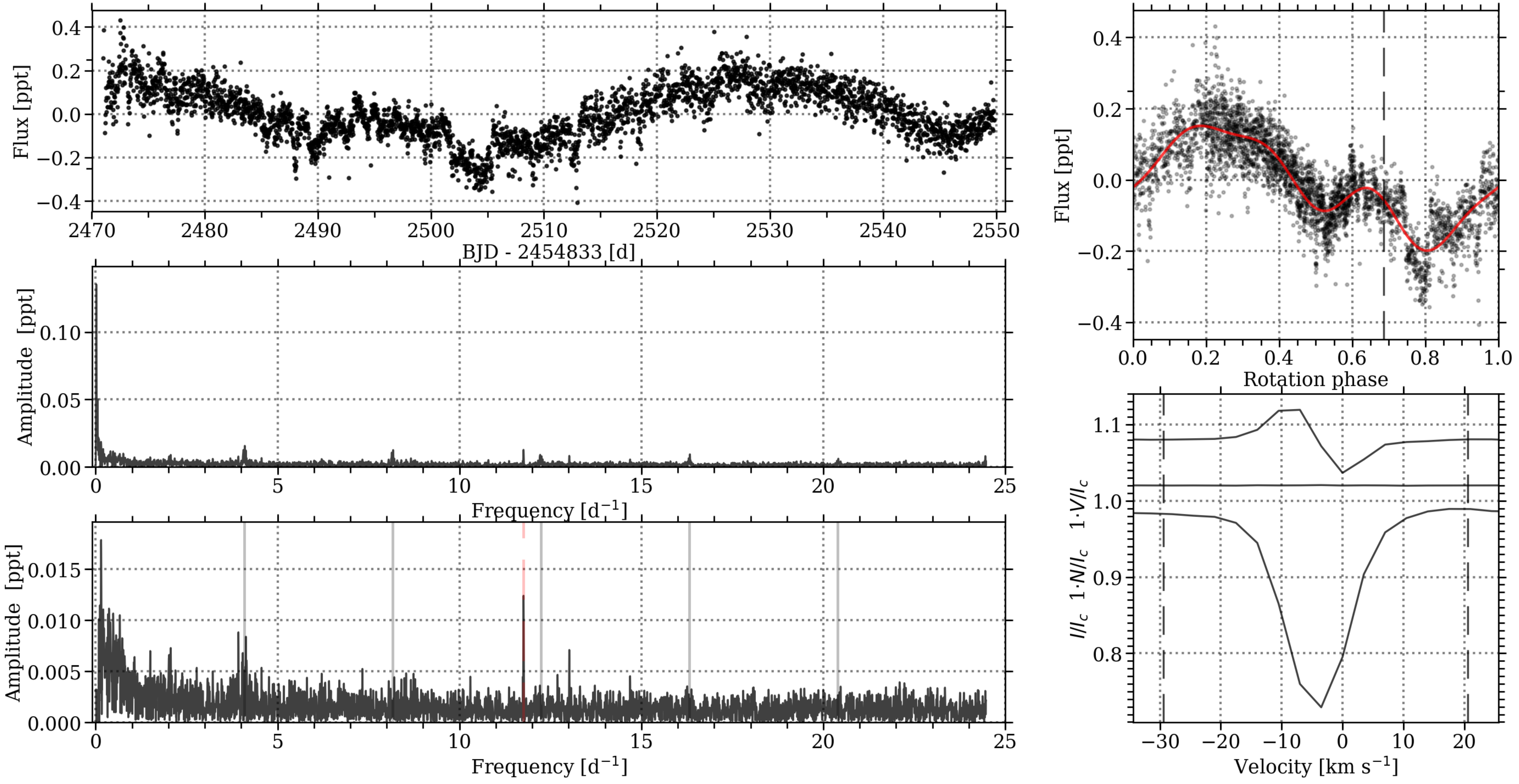}%
			\caption{Information related to HD\,177765, showing the K2 photometry, rotational modulation, and the LSD profiles of the ESPaDOnS data.  Same color coding was applied as Fig.\,\ref{fig:hd107000}.  The alias frequency of the presumed roAp pulsation was marked by the red dashed line in the \textit{bottom left} panel.}
			\label{fig:hd177765}
\end{figure*}

\bsp	
\label{lastpage}
\end{document}